\documentclass[a4paper,11pt]{article}
\pdfoutput=1
\usepackage[utf8]{inputenc}
\usepackage[pdftex]{graphicx}
\usepackage{amsfonts,amsmath,amssymb, amsthm}
\usepackage{color}
\usepackage[usenames,dvipsnames]{xcolor}
\usepackage{notoccite}
\usepackage{subcaption}
\usepackage{graphicx}
\usepackage{jheppub}
\usepackage{bbm}
\usepackage[ruled, noresetcount]{algorithm2e}
\usepackage{hyperref}
\usepackage{bm}

\newcommand{\tb}[1]{\textbf{#1}}

\theoremstyle{definition}
\newtheorem{definition}{Definition}

\title{Sparse SYK  and traversable wormholes}
\author[a]{Elena C\'aceres,}
\author[a]{Anderson Misobuchi,}
\author[a]{Rafael Pimentel}
\affiliation[a]{Theory Group, Department of Physics, University of Texas, Austin, TX 78712, USA}
\emailAdd{elenac@utexas.edu}
\emailAdd{anderson.misobuchi@utexas.edu}
\emailAdd{rafaelpimentel@utexas.edu}

\abstract{We investigate  two {\it sparse} Sachdev-Ye-Kitaev (SYK) systems coupled by a bilinear term as a holographic quantum mechanical description of an eternal traversable wormhole in the low temperature limit. Each SYK system consists of $N$ Majorana fermions coupled by random $q$-body interactions. The degree of sparseness is captured by a regular hypergraph in such a way that the Hamiltonian contains exactly $k\,N$ independent terms.  We improve on the theoretical understanding of the sparseness property by using known measures of hypergraph expansion.   We show that the sparse version of the two coupled SYK model is gapped with a ground state close to a thermofield double state. Using  Krylov subspace and parallelization techniques, we simulate the  system for  $q=4$ and $q=8.$ The sparsity of the model allows us to explore larger values of $N$ than the ones existing in the literature for the all-to-all SYK. We analyze in detail the two-point functions and the transmission amplitude of signals between the two systems. We identify a range of parameters where revivals {\it obey the scaling predicted by holography} and  signals can be interpreted as traversing the wormhole.}
\begin{document}

\maketitle


\section{Introduction}

The Sachdev-Ye-Kitaev (SYK) model \cite{KitaevTalks, Maldacena:2016hyu} is an exactly solvable quantum mechanical system of $N$ Majorana fermions coupled by a random all-to-all $q$-body interaction.  In the large $N$ and low-energy limit,  an approximate conformal symmetry emerges and the system  saturates the chaos bound \cite{Maldacena:2015waa}.  This unique combination of properties is also present in a large class of black holes  and suggests that  the SYK model is a holographic dual of a two-dimensional black hole \cite{Maldacena:2016upp}.  However, numerical computations in the all-to-all SYK  are resource intensive; the number of terms in the Hamiltonian grows as  $\sim N^q$. Recently, Xu, Susskind, Su and Swingle (XSSS) \cite{Xu:2020shn} proposed a \emph{sparse} version of the SYK model (see also \cite{Garcia-Garcia:2020cdo}). The advantage of the sparse SYK is that the interactions grow linearly with $N$, as opposed to $N^q$, while retaining properties of the original all-to-all SYK model such as fast scrambling and maximal chaos at low temperatures. The significant reduction in computational complexity of the sparse model opens up a new avenue where we can investigate properties of the SYK model at finite $N$ and study how a gravitational behavior emerges as $N$ increases. There is still much to be understood about the sparse SYK. For example, how sparse can the model be and still retain properties of the original SYK? Does it  reproduce all the physics of the all-to-all SYK or only some features? How large is the computational benefit of sparseness? The goal of this paper is to advance our understanding of these questions. 

A natural language to describe many-body interactions is provided by \emph{hypergraphs}. Quantum systems consisting of  two-body interactions can be described using \emph{graphs}; each degree of freedom lives on a vertex and the interaction is represented by edges connecting them appropriately. For many-body systems, graphs are not adequate anymore, we need to use \emph{hypergraphs}. The formal definition of hypergraphs is in terms of set theory but intuitively we can think of a hypergraph as a generalization of a graph where particles are identified as vertices and interaction terms in the Hamiltonian as \emph{hyperedges}  connecting \emph{more than two vertices}. To define a sparse model of a quantum many-body system  that still behaves as the complete model, we want to preserve enough hypergraph \emph{connectivity} while  keeping the smallest number of hyperedges possible. Hypergraphs that are highly connected but are sparse are known as expanders. A natural question arises: \emph{how sparse can a hypergraph be and still reproduce features of the all-to-all SYK?} To address this question we use a two prong approach: we  study formal properties of hypergraphs and also perform explicit calculations of physical quantities and compare the results to the all-to-all SYK. 

The central theme of the present work is to investigate to what extent sparse SYK models can be used to simulate features of an eternal traversable wormhole. Maldacena and Qi \cite{Maldacena:2018lmt} showed that two copies of SYK coupled via a bilinear interaction behave as an eternal traversable wormhole in the large $N$, low temperature and weak coupling limit. The wormhole has a global $AdS_2$ geometry where the two boundaries are causally connected and it remains traversable independently of time. This geometry can be obtained as a solution of Jackiw-Teitelboim (JT) gravity \cite{Jackiw:1984je, Teitelboim:1983ux, Almheiri:2014cka} by adding a time independent coupling between the two boundaries following the mechanism pioneered in \cite{Gao:2016bin}.\footnote{Alternatively, we can think of it as a solution of JT gravity coupled to matter fields but it requires matter that violates the Averaged Null Energy Condition (ANEC).} 
Some of the main features of the two coupled SYK system are that it has a ground state close to a thermofield double (TFD) with a temperature that depends on the strength of the coupling \cite{Maldacena:2018lmt, Garcia-Garcia:2019poj, Alet:2020ehp} and that there is  an energy gap above the ground state that plays an important role in the characterization of the dynamics of the model. In this work we demonstrate that these properties also appear in the sparse SYK model indicating that the two coupled  sparse SYK model is a good candidate for a  holographic dual of a traversable wormhole. 

To sharpen this connection we focus on the ability to send a signal through the wormhole. This phenomenon can be interpreted as a quantum teleportation protocol 
\cite{Maldacena:2017axo, Susskind:2017nto, Yoshida:2017non, Brown:2019hmk, Schuster:2021uvg, Nezami:2021yaq, Bao:2018msr, Gao:2019nyj} or as a generic feature of chaotic systems. 
In the context of quantum chaotic many-body systems, the addition of an instantaneous coupling between two quantum systems can be understood  as a \emph{revival} phenomena \cite{Gao:2018yzk}: a localized perturbation in one system  gets scrambled and reappears at later times on the other system due to the addition of the coupling. This effect is expected to be universal in quantum chaotic many-body systems, relying only on the chaotic nature of the system and the entanglement structure of its state. Since the interaction in the Maldacena and Qi model is time independent, the system of two coupled SYKs undergoes a series of revivals. From the gravity point of view, this is simply understood as the oscillatory motion of a particle that bounces back and forth between the two boundaries, traveling through the wormhole. This type of dynamics has been studied in detail in Ref.\,\cite{Plugge:2020wgc} and also at finite temperature in Ref.\,\cite{Qi:2020ian}. The frequency of the revival oscillations is controlled by the energy gap. 

In the system of two $q=4$ SYKs coupled via a bilinear term with coupling strength $\mu$, revival oscillations with a frequency $\propto \mu^{2/3}$ have been observed by solving the Schwinger-Dyson (SD) equations in the  $N \rightarrow \infty$ limit \cite{Plugge:2020wgc}. Remarkably, this nontrivial scaling is in perfect \emph{agreement with predictions from holography} \cite{Maldacena:2018lmt}. At finite $N$, however, this type of dynamics has not been found. Finite-size effects and the computational cost of the all-to-all SYK limit the state of the art simulations of two coupled SYKs to $2N=32$ Majorana fermions. In this work, we employ Krylov subspace methods (see Appendix \ref{app:numerical}) and massive parallelization to study a two coupled \emph{sparse} SYK system in detail for $2N=40$. Using these techniques, we simulate the sparse SYK system for $q=4$ and $q=8$, and show that the nontrivial frequency scaling, consistent with the causal structure of the eternal traversable wormhole with global $AdS_2$ geometry, is present for some range of values of the coupling $\mu$. Our numerical calculations make use of the \texttt{dynamite} package \cite{dynamite}, a Python wrapper for dynamical evolution based upon the SLEPc/PETSc libraries.

This paper is organized as follows. In Section \ref{sec:sparse-syk}, we introduce the relevant hypergraph definitions and review the sparse SYK model defined on a regular hypergraph. In Section \ref{sec:sparsity}, we characterize the amount of sparsity in the model using known measures of hypergraph expansion, namely the algebraic entropy, vertex expansion and spectral gap. In Section \ref{sec:twoSYK} we introduce the two coupled sparse SYK model and study its properties. In particular, we study the effect of sparsity on the system by focusing on the $q=4$ and $q=8$ body interaction cases.   In Section \ref{sec:transmission}, we present a diagnostic of signal transmission between the two sparse SYK system at both zero and finite temperature and we compare our results with the gravitational picture valid in the large $N$ limit. We conclude with some remarks and future directions in Section \ref{sec:discussion}.


\section{Preliminaries:  sparse SYK and hypergraphs} 
\label{sec:sparse-syk}

The Sachdev-Ye-Kitaev (SYK) model \cite{KitaevTalks} has been a successful toy model of lower dimensional quantum black holes. The model was proposed by Kitaev inspired by a spin quantum Heisenberg magnet with Gaussian distributed interactions previously studied by Sachdev and Ye \cite{Sachdev:1992fk}. Here, we will refer to the original SYK model as all-to-all SYK. The gravitational description is given by a nearly $AdS_2$ solution of two-dimensional JT gravity \cite{Almheiri:2014cka}. A variant of the SYK model, dubbed sparse SYK, was recently proposed as an effective theory for the all-to-all SYK with the advantage of allowing for more efficient computer simulation \cite{Xu:2020shn} (see also \cite{Garcia-Garcia:2020cdo}). In this section we will briefly review the sparse SYK model and some basic definitions pertaining hypergraphs which will be helpful to describe the structure of the interactions in the model.

\subsection{Review of the sparse SYK model}

The Hamiltonian of the sparse SYK model with $N$ Majorana fermions $\chi^j$, $j=1,\ldots, N$, is defined as
\begin{align} \label{eq:H_single}
	H & = i^{q/2}\sum_{j_1<\ldots <j_q}J_{j_1\ldots j_q}x_{j_1\ldots j_q}\chi^{j_1}\ldots\chi^{j_q}.
\end{align}
The parameters $x_{j_1...j_q}$ are either 0 or 1 and they can be defined in different ways leading to different sparse models. The couplings $J_{j_1...j_q}$ are drawn from a Gaussian distribution with zero mean and variance given by
\begin{equation} 
	\langle \left(J_{j_1\ldots j_q}\right)^2\rangle = \frac{(q-1)!J^2}{pN^{q-1}},
\end{equation}
where $p$ is the fraction of terms in the Hamiltonian, i.e., the number of the terms such that $x_{j_1...j_q}=1$ divided by $\binom{N}{q}$, which is the number of terms in the all-to-all version. The parameter $J$ has dimensions of energy and sets the energy scale of the theory. It will be convenient to define the parameter
\begin{equation} \label{eq:k}
    k \equiv \frac{p}{N}\binom{N}{q},
\end{equation}
which is such that the Hamiltonian is a sum of $kN$ independent terms. Due to the random nature of the couplings, physical observables are obtained after performing an average over different disorder realizations of the system. If we choose all $x_{j_1\ldots j_q}$ to be 1, we recover the original all-to-all SYK Hamiltonian
\begin{equation} \label{eq:hamiltonian_all}
    H_\text{all-to-all} = \sum_{1\leq j_1<\ldots<j_q\leq N}J_{j_1\ldots j_q}\chi^{j_1}\ldots\chi^{j_q}.
\end{equation}
In this case we have $p=1$ and we also recover the variance of the all-to-all SYK model
\begin{equation}
	\langle \left(J_{j_1\ldots j_q}\right)^2\rangle_\text{all-to-all} = \frac{(q-1)!J^2}{N^{q-1}}.
\end{equation}
One possible implementation of sparseness in the SYK model consists of taking $x_{ijkl}=1$ with probability $p$ and $x_{ijkl}=0$ with probability $1-p$. The system becomes more sparse as $p\to 0$ and we recover the all-to-all SYK when $p=1$. This procedure, known as \emph{random pruning}, is perhaps the simplest to implement. However, it makes necessary to take an additional average since $x_{ijkl}$ are treated as random variables and can potentially lead to disconnected clusters of Majorana fermions that do not interact with the rest of the system \cite{Garcia-Garcia:2020cdo}. A more powerful and general approach to characterize the sparseness in the SYK model is to use hypergraphs. 

\subsection{Hamiltonian interactions and sparsity from hypergraphs}

The structure of Hamiltonian interactions of a quantum many-body system can be viewed as a hypergraph. Before plunging into hypergraphs and its properties let us recall that an (undirected) graph $G$ can be defined as a tuple $G=(V,E)$ where $V$ is a finite set whose elements are called vertices and $E$ is a set of pairs of vertices. Every element $e \in E$ is encoded as a set of exactly 2 vertices, i.e., $|e| = 2$.  Because of that, graphs are good tools to model problems where the relationship between objects comes in pairs, such as in links of computer networks, couplings in Ising or general 2-local spin chains, etc. However, interactions or relationships between more than two elements cannot be described in terms of graphs. A hypergraph is then a natural generalization of a graph where a hyperedge can contain any number of vertices. Thus, hypergraphs become necessary to describe the SYK model with $q>2$ body interactions.

The definitions below will help establish our notation and facilitate the discussion in the following sections \cite{ book:Bretto, Dumitriu_2019}: 

\begin{definition}
    A \tb{hypergraph} $H=(V,E)$ consists of a set $V$ of vertices and a set $E$ of hyperedges such that each hyperedge is a nonempty subset of $V$. \end{definition}
\begin{definition}
    A hypergraph $H$ is $\bm{s}$\tb{-uniform} for an integer $s\ge 2$ if every hyperedge  $e \in E $ contains exactly $s$ vertices. \end{definition}
\begin{definition}
    The \tb{degree} of the vertex $i$, denoted $\text{deg}(i)$, is the number of all hyperedges incident to $i$.\end{definition}
\begin{definition}
    A hypergraph is $\bm{r}$\tb{-regular} if all of its vertices have degree $r$.\end{definition}
\begin{definition}
    A hypergraph is $\bm{(r, s)}$\tb{-regular} if it is both $r$-regular and $s$-uniform.\end{definition}
\begin{definition} 
    A $s$-uniform hypergraph is \tb{complete} if  the edge set is the set of all $s$-element subsets of vertices.\end{definition}

Let us illustrate these definitions for the all-to-all SYK with $N$ Majoranas and $q$-body fermion interactions. Each Majorana is identified with a vertex, and each term in the Hamiltonian is a hyperedge with $q$ vertices, meaning that the corresponding hypergraph is $q$-uniform. The number of terms in the Hamiltonian, i.e., the number of hyperedges is
\begin{equation}
    L_\text{all-to-all}=\binom{N}{q}.
\end{equation}
The fact that the system is `all-to-all' translates into the hypergraph being complete. That is, the edge set has cardinality $|E|= \binom{N}{q}$ which is the number of all possible $q$-element subsets of $N$. On the other hand, the sparse SYK hypergraph is still $q$-uniform but the Hamiltonian has fewer terms than the all-to-all and clearly the associated hypergraph is no longer complete. We can write the number of terms as
\begin{equation}
    L_\text{sparse}= p\, \binom{N}{q},
\end{equation}
where $0 < p \leq 1$. In terms of the parameter $k$ defined in \eqref{eq:k}, this can be rewritten as
\begin{equation}
    L_\text{sparse}= k\, N.
\end{equation}

\begin{figure}[t]
    \centering
    \begin{subfigure}[b]{0.45\textwidth}
        \centering
        \includegraphics[width=\textwidth]{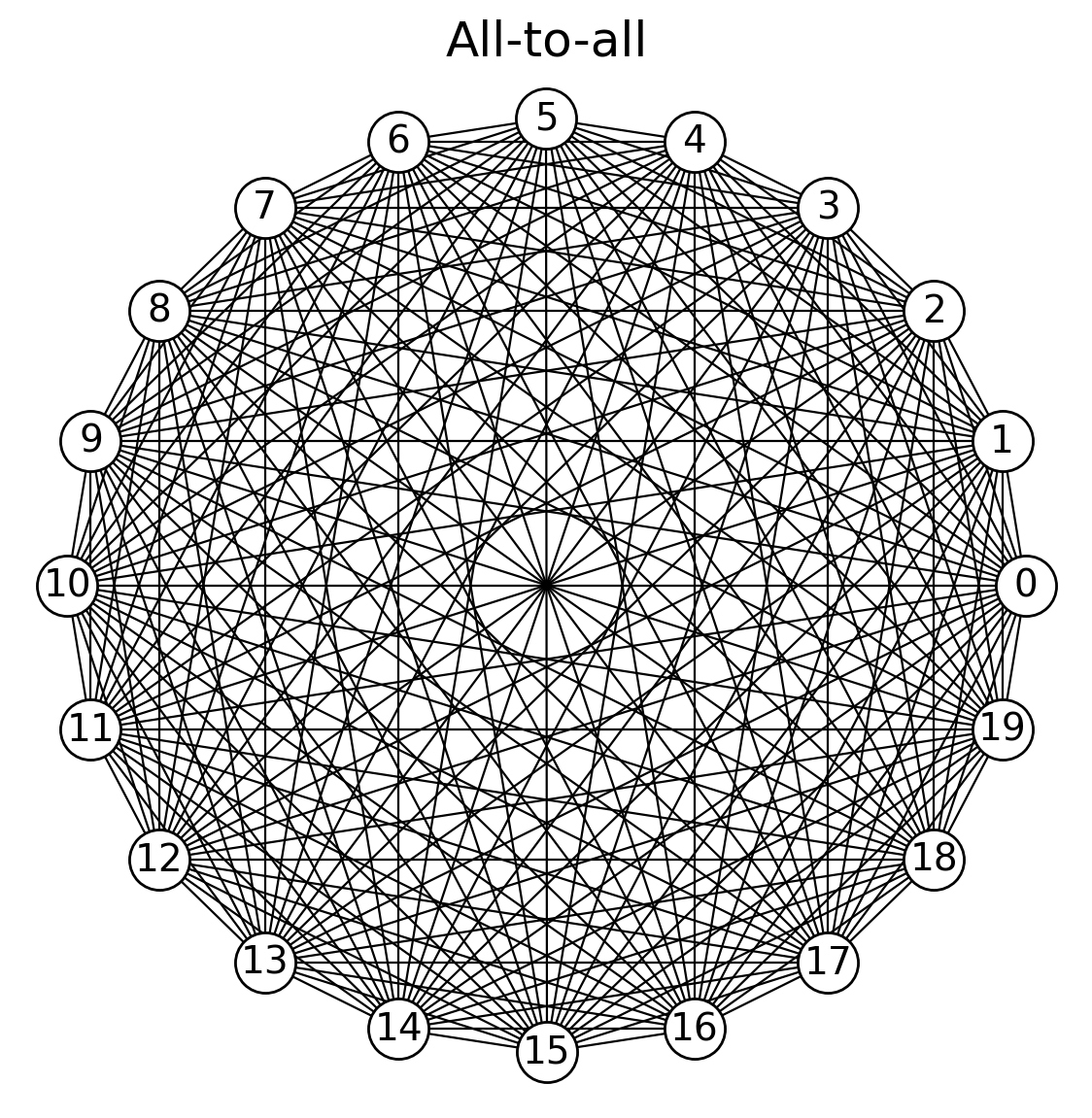}
    \end{subfigure}
    \begin{subfigure}[b]{0.45\textwidth}
        \centering
        \includegraphics[width=\textwidth]{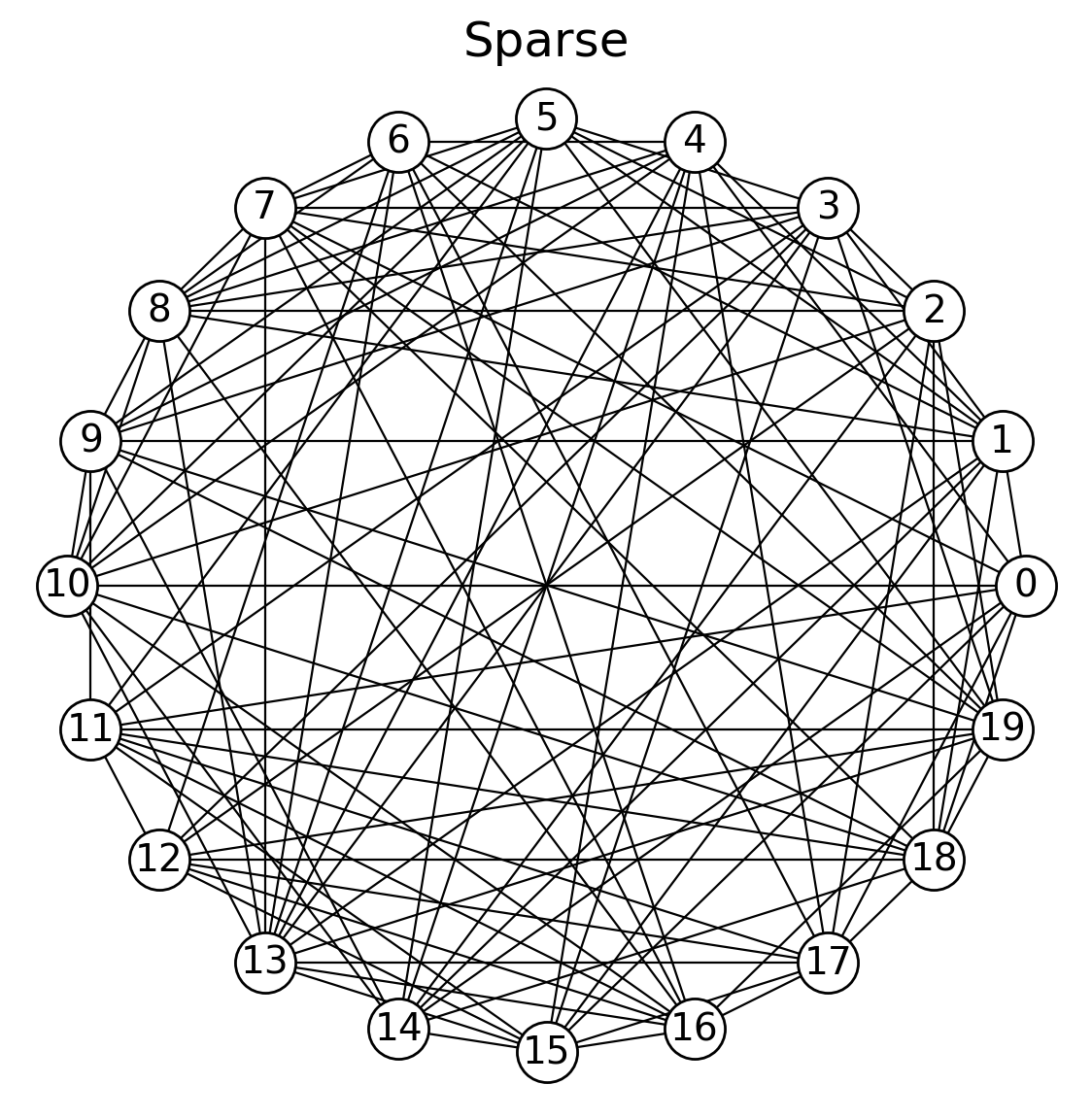}
    \end{subfigure}
    \caption{Hypergraphs associated to the Hamiltonians for the all-to-all SYK (Left) and sparse SYK from a $(k\,q, q)$-regular hypergraph (Right). The parameters used are $N=20$, $q=4$ and $k=1$. For drawing purposes, each hyperedge is replaced by a complete set of edges.}
    \label{fig:graph}
\end{figure}

An important property of $(r,s)$-regular hypergraphs is that they are expected to be \emph{expanders} \cite{Dumitriu_2019, Li_Mohar}. Expanders are sparse but still have good connectivity, in a sense that we will make precise in the next section. Thus, we want to define the sparse SYK model over a regular, uniform hypergraph. More precisely, the sparse SYK Hamiltonian will be defined over a $(k\,q, q)$-regular hypergraph. It is $q$-uniform because all terms in the Hamiltonian still involve only $q$-body interactions, and the fact that it is $k\,q$-regular follows from a simple counting argument.\footnote{Recall that there are $k\,N$ hyperedges and each hyperedge contains $q$ vertices. Thus, the full list of hyperedges contains $k\,N\,q$ vertices. By imposing the regularity condition, each vertex must appear exactly $k\,q$ times on the list of hyperedges and we conclude that the degree of each vertex should be $r=k\,q$.} We illustrate the difference between the all-to-all and the sparse hypergraphs in Fig.\,\ref{fig:graph}.

Summarizing, the sparse SYK Hamiltonian can be represented as 
\begin{equation}
	H = \sum_{i=1}^{L_\text{sparse}}H_{\{v_i\}},
\end{equation}
where $H_{\{v_i\}}$ acts on a set of vertices $\{v_i\}$. The hypergraph associated to this Hamiltonian is $q$-uniform and $r$-regular with $r= k\,  q$. In other words, it is a   ${(k\, q, q)}$-regular hypergraph. The parameter $k$ quantifies the degree of sparseness in the Hamiltonian. In Ref.\,\cite{Garcia-Garcia:2020cdo} the authors described an efficient algorithm, Algorithm \ref{alg:regularhyper}, based on the so called `pairing model' for regular graphs \cite{Wormald}, that generates a regular hypergraph randomly. The same algorithm will be employed in our simulations. In the next section we turn our attention to the question of how to quantify the connectivity of a hypergraph.

\IncMargin{1em}
\begin{algorithm}[t]

\KwResult{If at the end $\ell$ is empty then $E$ gives a list of hyperedges for a $(k\,q, q)$-regular hypergraph}
    Initialize $\ell=\{1,1,1, 2,2,2,\ldots, N,N,N\}$ where each integer $1,\ldots,N$ appears exactly $kq$ times\;
    Initialize empty list to store hyperedges $E = \{\}$\;
    \For{$n=1$, $n<2N$ (or more)}{
        Sample $q$ elements from $\ell$ (i.e., a hyperedge)\;
        \If{sample has no duplicates and it is not already in $E$}{
            Move sample to $E$\;
            Remove sample from $\ell$\;
            }
        $n\gets n+1$
    }
    \caption{Generate a $(k\,q,q)$-regular hypergraph }\label{alg:regularhyper}
\end{algorithm}
\DecMargin{1em}

\section{How much sparseness?}
\label{sec:sparsity}

As previously mentioned, the advantage of the sparse SYK model is that due to its high connectivity it has the same features as the all-to-all SYK -- such as quantum chaos and scrambling -- but the sparsity makes it less computational demanding. A central question is: how much sparsity are we allowed to have and at the same time maintain these features?  This is one of the questions we aim to answer in this paper. Intuitively, we want the hypergraph to have a small number of hyperedges but in a way that it maintains `enough' connectivity. Quantifying connectivity is a central question in the theory of sparse hypergraphs. There is a class of hypergraphs that are sparse and at the same time have strong connectivity properties; they are known as \emph{expanders}. Random regular hypergraphs are believed to be  expanders \cite{Dumitriu_2019,Li_Mohar}. That is the main motivation to define the sparseness in the Hamiltonian in such a way that the associated hypergraph is $(k\,q, q)$-regular.

In order to get intuition for what is an expander hypergraph, let us first review the concept of an expander graph. The connectivity of a graph can be quantified using measures of edge expansion, vertex expansion or spectral expansion. 
A standard measure of edge expansion is the Cheeger constant. Let $G$ be a $k$-regular  graph (with $k\le 2$) with $n$ vertices, and $S$ a subset of $G$ of cardinality $|S|$. The Cheeger constant is defined as  
\begin{equation}\label{eq:cheeger}
    h(G)=\underset{0\le |S|\le \frac{n}{2}}{\textrm{min}}\frac{|\partial S|}{|S|},
\end{equation}
where $\partial S$ is the set of edges running from a vertex in $S$ to  a vertex outside of $S$, and $|\partial S|$ is its cardinality. If  $|S|\le \frac{n}{2}$, then $|\partial S|\ge h |S|$ so  that  if  $h$ is large,  then every  subset $S$ has many neighbors outside $S$ and $G$ is said to be an expander. We can also measure the expansion of a graph by looking at the vertex expansion. In this case, we are interested in comparing the number of vertices in a neighborhood of a subset to the vertices in the subset. The third measure of expansion, the spectral gap, refers to the eigenvalues of the adjacency matrix. The adjacency matrix $A$ is a $n\times n$ matrix whose elements are either $1$ or $0$ depending on if there is an edge that connects the two vertices or not. Since $A$ is symmetric, the eigenvalues are guaranteed to be real. The spectral gap is defined as the difference between the largest and the second largest eigenvalues. In graph theory, it has been shown that these three different measures of expansion -- edge, vertex, spectral  gap -- are all related to the eigenvalues of the adjacency matrix; they all provide equivalent measures of the connectivity of a graph.

On the other hand, \emph{hypergraph expansion} is much less understood. Unlike for graphs, it is not clear which operator associated to the hypergraph captures information about the connectivity. One approach is to generalize the definition of an adjacency matrix for hypergraphs \cite{Feng_Li}. Other authors suggest to define an appropriate tensor \cite{Friedman_Wigderson}. A third approach relies on Laplacian matrices defined through  higher order random walks \cite{Lu_Peng}. In this paper we will follow the adjacency matrix approach of \cite{Feng_Li} and use some recent results showing that vertex expansion is controlled by the second largest eigenvalue of the adjacency matrix and that for a random regular hypergraph there is always a spectral gap \cite{Dumitriu_2019}. 

Let us begin by defining the adjacency matrix of a hypergraph. 

\begin{definition} 
    The adjacency matrix associated to a hypergraph with $n$ vertices is the $n\times n$ matrix whose elements are
    \begin{equation}
        A_{ij}=\begin{cases}
            \begin{aligned}
                &\textrm{\# of hyperedges containing vertices $i$ and $j$}
            \end{aligned} 
            &\textrm{if } \,i\neq j \\
        \qquad \qquad 0 & \textrm{if } \, i=j
        \end{cases}.
    \end{equation}

\end{definition}

In this section we will discuss  a measure of hypergraph vertex expansion and apply it to $(k\,q, q)$-regular hypergraphs that describe the sparse SYK Hamiltonian. We will also discuss the algebraic entropy that measures how much information is contained in a hypergraph. These two measures will guide us to choose a sparsity that maximizes numerical efficiency and at the same time preserves properties of the all-to-all SYK. 

\subsection{Algebraic hypergraph entropy}

Hypergraphs are used in many areas, among them computer science, biology, neural networks, engineering, etc. It is important to quantify the amount of information contained in a hypergraph. A hypergraph is a combinatorial object that can be very complicated, so the notion of entropy can also serve to assess this complexity. The definition we present here to quantify this information is the \emph{algebraic hypergraph entropy} $I(H)$ \cite{book:Bretto}. 

Consider a hypergraph $H=(V,E)$ and its adjacency matrix $A(H)$ with elements  $A_{ij}$. We define the degree matrix as
\begin{equation}\label{eq:define_D}
    D = \text{diag}( d_1, d_2, \ldots, d_N)), \qquad d_i=\sum_{j \in V} A_{i j}.
\end{equation}
The Laplacian of $H$ is defined as 
\begin{equation}
    L(H)=D-A(H).
\end{equation}
The adjacency matrix $A(H)$ is symmetric and $A_{ii}=0$ by definition.
We also define a rescaled Laplacian
\begin{equation}\label{eq:rescaled_lap}
    L'(H) = \frac{L(H)}{\text{Tr}\,D}.
\end{equation}
The eigenvalues of the rescaled Laplacian, denoted by $\nu_i$, satisfy
\begin{equation}
    0\le \nu_i\le \, 1 \quad\quad \text{and} \quad\quad \sum_{i=1}^{N} \nu_i =1.
\end{equation}
We can now define the algebraic hypergraph entropy as 
\begin{equation}\label{eq:algebraic_entropy}
 I(H) = \,-\, \sum_{i=1}^N \, \nu_i \log \nu_i.
\end{equation}
Our goal is to explore different measures of how similar the sparse SYK is compared to the all-to-all version. As a first step, we calculate the algebraic entropy for the all-to-all-SYK, whose hypergraph is complete. Its adjacency matrix has elements
\begin{equation} \label{eq:A_all}
    A_{ij}=\binom{N-2}{q-2}\equiv w \quad \textrm{for } i\neq j, \quad \textrm{and }\quad A_{ii}=0.
\end{equation}
From \eqref{eq:define_D}, we immediately obtain that
\begin{align}
    & D = \text{diag}(d_1, \ldots, d_N) = w\, (N-1) \, \mathbbm{1}_{N\times N}.
\end{align}
The corresponding rescaled Laplacian \eqref{eq:rescaled_lap} has components
\begin{equation}
    L_{ij}'=- \frac{1}{N(N-1)} \quad \textrm{for } i\ne j, \quad \textrm{and}\quad L_{ii}'=0.
\end{equation}
The eigenvalues of $L'$ turn out to be $\nu_1 = 0 $ and $\nu_2,\ldots,\nu_N=\frac{1}{N-1}$, so that the algebraic entropy for the all-to-all SYK is
\begin{equation}
    I(H_\text{all-to-all})=-\sum_i \nu_i \log \nu_i = \log(N-1).
\end{equation}

\begin{figure}[t]
    \centering
    \includegraphics[width=0.6\linewidth]{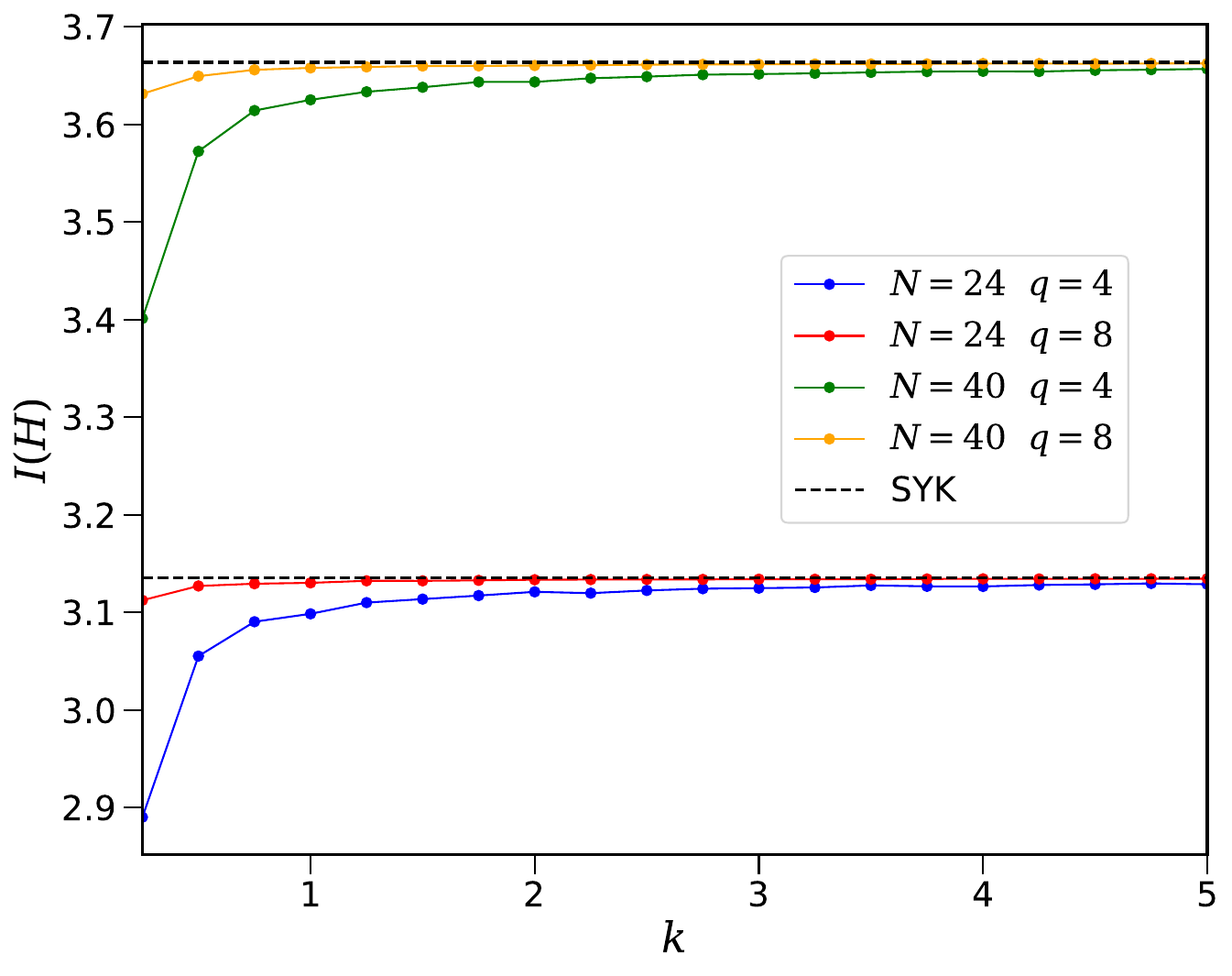}
    \caption{Algebraic hypergraph entropy $I(H)$ as a function of the sparsity parameter $k$ for a $(kq,q)$-regular hypergraph generated randomly using Algorithm \ref{alg:regularhyper}. The entropy is obtained from the eigenvalues of the rescaled Laplacian matrix \eqref{eq:rescaled_lap}.
    \label{fig:entropy}}
\end{figure}

On the other hand, the algebraic entropy for a sparse SYK defined over a $(k\,q, q)$-regular hypergraph is determined numerically. In Fig.\,\ref{fig:entropy} we present the algebraic hypergraph entropy as a function of the sparsity parameter $k$ and for the all-to-all SYK. Clearly, the $q=8$ result approaches the all-to-all entropy value for smaller $k$ as compared to $q=4$. We will see that this is a pattern common in all the measures of connectivity. 

\subsection{Spectral gap and vertex expansion}

In graph theory,  the difference between the two largest eigenvalues of the Laplacian \eqref{eq:rescaled_lap} associated to a graph $G$, also known as spectral gap, plays an important role in the discussion of expander graphs. Generically, the Laplacian is a scaled and shifted version of the adjacency matrix $A(G)$. For the special case of regular graphs, the Laplacian and the adjacency matrix are related in such a way that  studying the eigenvalues of the adjacency or of the Laplacian is equivalent. Furthermore, if we denote $\lambda_\text{max}\ge \lambda_2\ge \dots \ge\lambda_n$ the eigenvalues of $A(G)$, the largest eigenvalue of a $d$-regular graph always satisfies $\lambda_\text{max}=d$. Thus, in the case of regular graphs  all the relevant information is contained in the second largest eigenvalue
\begin{equation}\label{eq:lambda_second_def}
    \lambda\equiv\textrm{max}\{\lambda_2,|\lambda_n|\}.
\end{equation}

Since all the important information is contained in $\lambda$, this quantity is commonly referred to, as an abuse of language, as the \emph{spectral gap}. Even though vertex expansion and spectral gap are well understood for $d$-regular \emph{graphs}, the situation is very different for hypergraphs. Just very recently \cite{Dumitriu_2019}, the authors proved that uniformly  random  $(r,s)$-regular  hypergraphs  always have a spectral  gap.  The largest eigenvalue of the adjancency matrix for  $(r,s)$-regular hypergraphs is  $\lambda_\text{max} = r (s-1)$. Thus, hyperedge and vertex expansion are controlled by the second largest eigenvalue of the adjacency matrix. Therefore, we can study the spectral gap of the sparse SYK model and compare it with the all-to-all SYK to obtain another measure of how good an expander a given random $(r,s)$-regular hypergraph is. 

\begin{figure}
    \centering
    \includegraphics[width=0.48\linewidth]{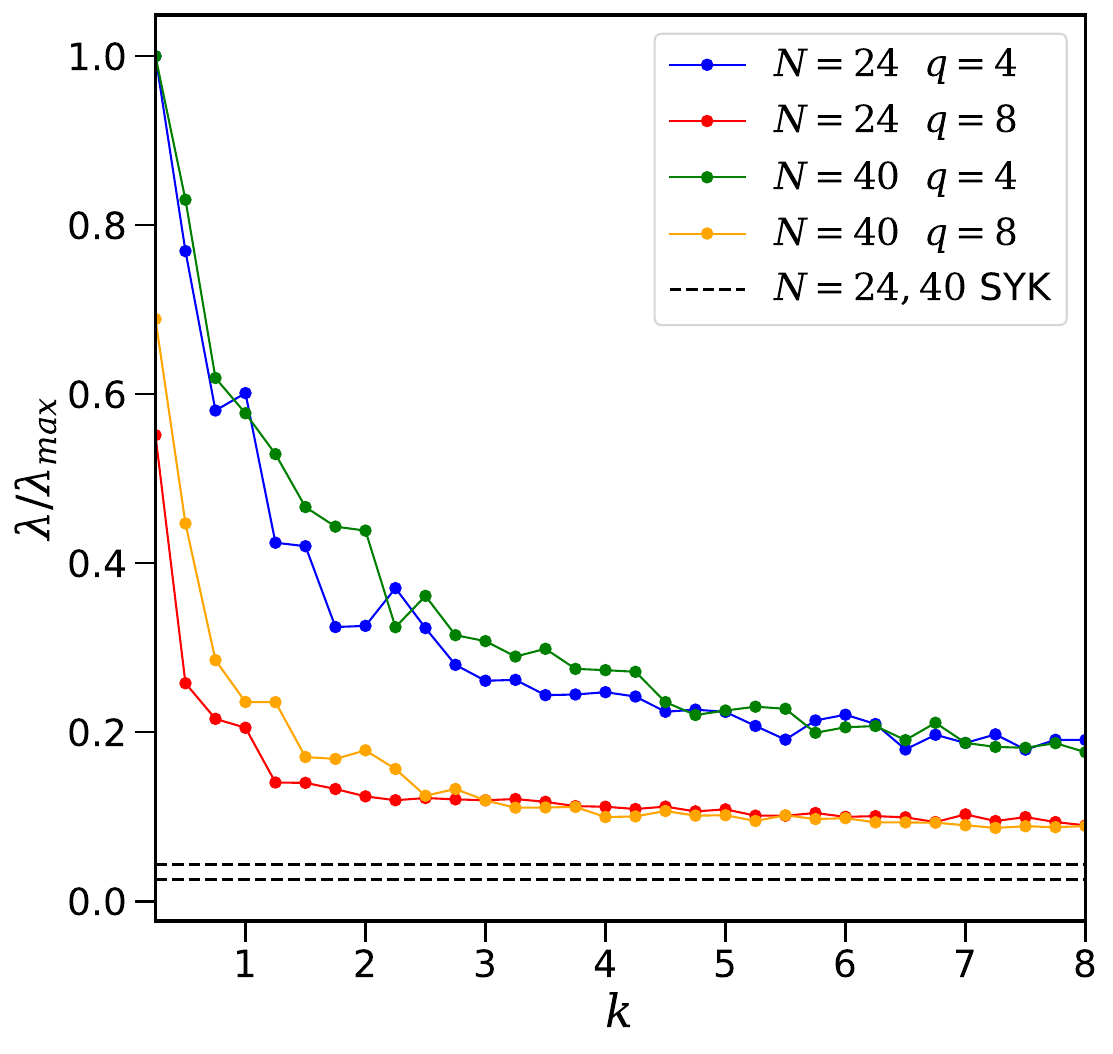}
    \includegraphics[width=0.48\linewidth]{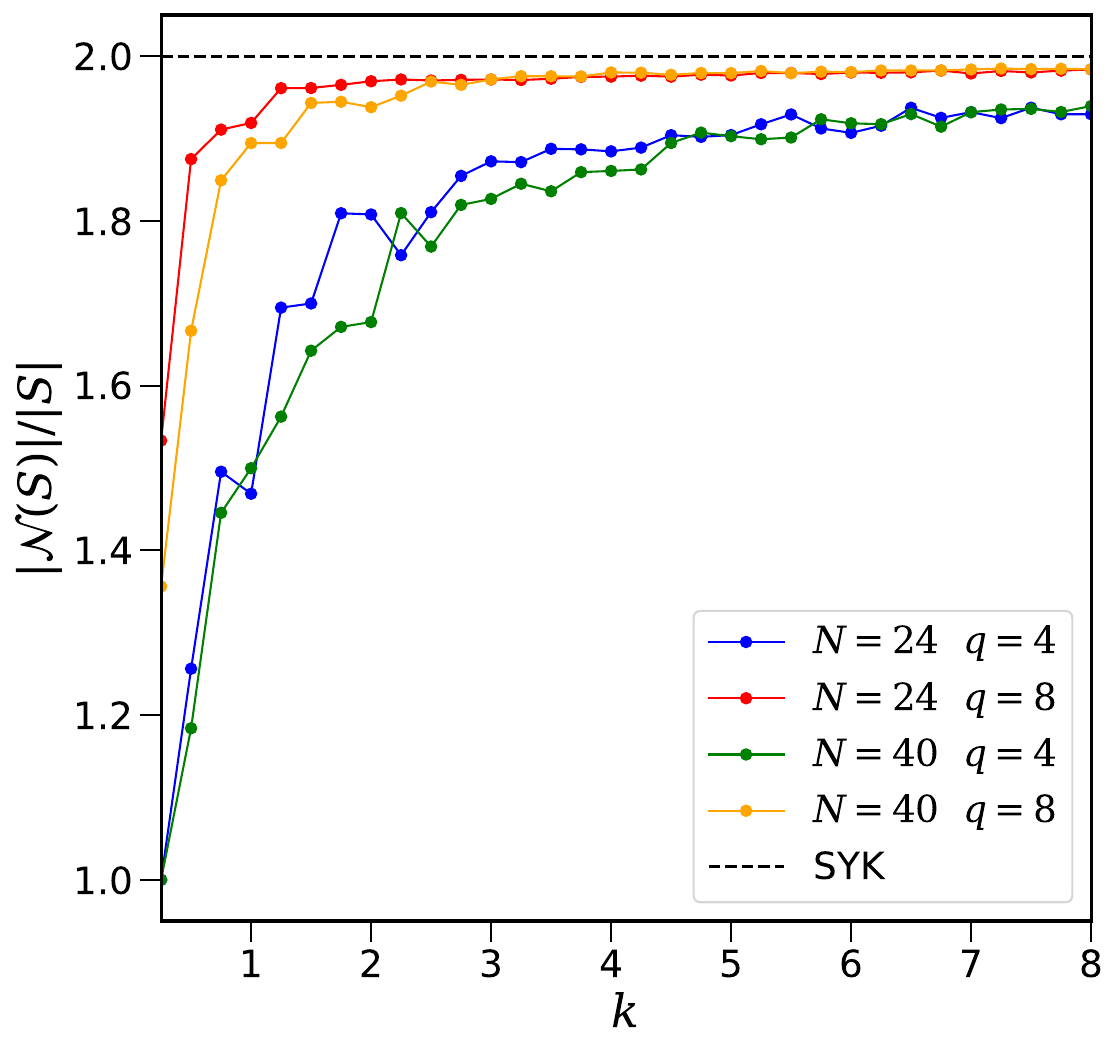}
    \caption{\textit{Left:} Spectral gap normalized by the largest eigenvalue of the adjacency matrix. \textit{Right:} Bound on vertex expansion \eqref{eq:vertex_half} for subsets $S\subset V$ smaller than half the size of the vertex set $V$. The dashed lines are the values for the all-to-all SYK.}
    \label{fig:vertex_gap}
    \end{figure}

We note that the largest eigenvalue of the adjacency matrix \eqref{eq:A_all} associated to the all-to-all Hamiltonian \eqref{eq:hamiltonian_all} is $\tilde{\lambda}_\text{max} = (N-1)\, w$ and all the other eigenvalues are equal $\tilde{\lambda}_2,\ldots,\tilde{\lambda}_n = -w$, which gives the spectral gap
\begin{equation}
    \tilde{\lambda}=w,
\end{equation}
where $w$ was defined in \eqref{eq:A_all}. Note that the ratio $\tilde{\lambda}/\tilde{\lambda}_\text{max} $ only depends on $N$ via
\begin{equation}
   \frac{ \tilde{\lambda}}{\tilde{\lambda}_\text{max}}=\frac{1}{N-1}.
\end{equation}
For the sparse SYK represented by a $(k\,q, q)$-regular hypergraph, its spectral gap $\lambda_\text{sparse}$ is evaluated numerically for each value of $k$. The results are illustrated in Fig.\,\ref{fig:vertex_gap}. Again, we see that $q=8$ requires a smaller value of $k$ to achieve a spectral gap closer to the all-to-all SYK. 

Another standard measure of expansion is \emph{vertex expansion}. Intuitively, vertex expansion is the idea  that small sets of vertices in a hypergraph  have a large numbers of neighbors. There is rich literature  regarding graphs vertex expansion. For example, Tanner's Theorem  \cite{Tanner} provides a lower bound on the size of a neighbourhood $\mathcal{N}(S)$ of a subset $S \subset V$. Before proceeding, let us define the neighborhood of $S$.

\begin{definition}
Consider a subset $S\subset V$. We define its \textit{neighborhood} as the set
\begin{equation}
   \mathcal{N}(S):=\{i: \exists\, j \in S\, \text{ such that }\{i,j\}\subseteq e\, \text{ for some } e \in E\}.
\end{equation}
\end{definition}

In \cite{Dumitriu_2019} the authors proved a theorem that bounds the vertex expansion of a random regular hypergraph. Consider a $(r,s)$-regular hypergraph $H=(V,E)$ with adjacency matrix $A(H)$. Let $\lambda:=\textrm{max}\{\lambda_2 (A), |\lambda_n(A)|\}$ where $\lambda_2(A)$ is the second largest eigenvalue of $A$ and $\lambda_n(A)$ denotes all the other eigenvalues. In Ref.\,\cite{Dumitriu_2019}, the authors showed that for any $S\subset V$
\begin{equation}
    \frac{|\mathcal{N}(S)|}{|S|}\geq \left[1-\left(1-\frac{\lambda^2}{r^2(s-1)^2}\right)\left(1-\frac{|S|}{n}\right)\right]^{-1}.
\end{equation}
We consider subsets $S$ that are at most half of the size of the set $V$, then  $\left(1-\frac{|S|}{n}\right) \ge 1/2$ and we have 
\begin{equation} \label{eq:vertex_half}
    \frac{|\mathcal{N}(S)|}{|S|}\geq \left[1-\frac{1}{2}\left(1-\frac{\lambda^2}{r^2(s-1)^2}\right)\right]^{-1}.
\end{equation}
This lower bound on vertex expansion is plotted in Fig.\,\ref{fig:vertex_gap}, right panel,  together with the all-to-all value. We observe that for $q=8$  the bound is always larger and closer to the all-to-all value. Furthermore,  if $q=4$ we need a larger value of $k$, as compared to $q=8$, to achieve the same level of vertex expansion. This implies that when simulating the sparse SYK we can expect that given a value of $N,$ the optimal value of $k$ will be $q$ dependent.
    

\section{Two coupled sparse SYK model}
\label{sec:twoSYK}

\subsection{Definitions of the model}
The traversable wormhole mechanism of Gao, Jafferis, and Wall \cite{Gao:2016bin} relies on turning on an interaction between the two asymptotic boundaries of a non-traversable wormhole for some amount of time, which renders the wormhole traversable for some finite time interval. Based on this idea, Maldacena and Qi \cite{Maldacena:2018lmt} constructed an eternal traversable wormhole with a global $AdS_2$ geometry that is traversable for any time. By considering a system made of two copies of the SYK model with a bilinear coupling between them, it has been found that the two coupled system behaves as the eternal traversable wormhole in the large $N$, low-energy, and small coupling limit. Here, we consider the sparse version of the two coupled SYK system, whose Hamiltonian is given by
\begin{equation} \label{eq:coupledH}
    H = H_{L} + H_{R} + H_\text{int}, \qquad H_\text{int} = i\mu \sum_{j}\chi_{L}^j\chi_{R}^j,
\end{equation}
where $H_{L}$ and $H_{R}$ are the Hamiltonians for two identical sparse SYK systems coupled by the bilinear interaction $H_\text{int}$ whose strength is controlled by the parameter $\mu$. We assume that each copy, which we refer to as $L$ (Left) and $R$ (Right), has $N$ Majorana fermions denoted by $\chi_a^j$, with $a=L,R$ and $j=1,\ldots, N$, and are normalized such that $\{\chi^i_a,\chi^j_b\} = \delta_{ab}\delta^{ij}$.  We recall that the Hamiltonian for a single sparse SYK was defined as
\begin{equation} \label{eq:H_coupled}
    H_a = i^{q/2} \sum_{1\leq j_1<\ldots<j_q\leq N} J_{j_1\ldots j_q}^a  x_{j_1\ldots j_q} \chi_a^{j_1}\ldots\chi_a^{j_q},
\end{equation}
where the coefficients $x_{j_1\ldots j_q}$ are either 0 or 1 according to the hypergraph that characterizes the sparsity in the model. We will restrict ourselves to the class of hypergraphs that are $(k\,q, q)$-regular (see definitions in Section \ref{sec:sparse-syk}), meaning that the associated Hamiltonian is a sum of exactly $k\,N$ terms and each Majorana appears in exactly $k\,q$ terms in the Hamiltonian. The random couplings $J^a_{j_1\ldots j_q}$ are sampled from a Gaussian distribution of zero mean and variance
\begin{equation}
    \langle (J^a_{j_1\ldots j_q})^2\rangle = \frac{2^{q-1}\mathcal{J}^2(q-1)!}{p q N^{q-1}} = \frac{J^2(q-1)!}{p N^{q-1}}.
\end{equation}
Note that the variance is rescaled by a factor $p=kN/\binom{N}{q}$ compared to the all-to-all SYK model. The parameter $J$ sets the energy scale of the system and we will work with the convention $J=1$.\footnote{Note that our convention is the same as in Refs. \cite{Xu:2020shn, Garcia-Garcia:2020cdo, Plugge:2020wgc, Garcia-Garcia:2019poj}, whereas Refs. \cite{Maldacena:2018lmt, Qi:2020ian, Alet:2020ehp} use $\mathcal{J}$ to define the energy scale of the system.} It is important to notice that, even though the Left and Right couplings are random, they are not independent. They are related by
\begin{equation}
    J_{j_1\ldots j_q}^L = i^{q}J_{j_1\ldots j_q}^R.
\end{equation}
This correlation is necessary to increase the information transfer between the two systems and observe revivals \cite{Maldacena:2018lmt, Plugge:2020wgc}.\footnote{A study of the two coupled SYK with imbalanced interactions have been considered in Ref.\,\cite{Haenel:2021fye}. Similar physics of the two coupled SYK has also been shown to arise by considering a single PT symmetric SYK \cite{Garcia-Garcia:2021elz}.} 

In our numerical simulations, we will make use of Krylov subspace methods and the Jordan-Wigner transformation (see Appendix \ref{app:numerical}) that maps the two coupled system with $2N$ Majoranas into a system of $N$ qubits.

\subsection{Real time Green's function}

We begin by computing the real time retarded Green's function for the two coupled sparse SYK model and we examine the effect of the sparsity characterized by the parameter $k$ defined above. Our goal is to estimate an `optimal sparsity', given by the smallest value of $k$ that still captures the same properties as the all-to-all SYK to a good approximation. Previous analysis \cite{Xu:2020shn, Garcia-Garcia:2020cdo} have indicated that $k\sim \mathcal{O}(1)$ is enough to reproduce features of SYK such as the chaotic behavior. Based on our refined analysis in Section \ref{sec:sparsity}, we expect that the optimal sparsity for the $q=8$ body system can be achieved using a smaller value of $k$ (more sparse) compared to the $q=4$ case, and here we verify this is indeed the case within the range of parameters we are able to explore. Once we have determined the optimal sparsity, we will move to the simulation of the two coupled system by fixing the optimal sparsity we have found, whose results we expect to also hold for the corresponding all-to-all SYK model.

The real time (retarded) Green's function $G_{ab}(t)$, averaged over all fermionic modes, is defined as
\begin{align} \label{eq:green-retarded}
    G_{ab}(t) & = \frac{1}{N} \sum_j 2\text{Re}\langle\chi_a^j(t)\chi_b^j(0)\rangle, \qquad a,b = L,R.
\end{align}
In the above expression, $\langle\ldots\rangle = \frac{1}{Z}\text{Tr}[\ldots e^{-\beta H}]$ denotes the thermal average at inverse temperature $\beta$ with normalization $Z=\text{Tr}[e^{-\beta H}]$, and it is implicitly understood that besides the thermal average there is also an average over quenched disorder for the different realizations of the random couplings $J_{j_1\ldots j_q}$ in the Hamiltonian. 

In the low-energy, weak coupling limit $\mu\ll J$, the large $N$ physics of the two coupled SYK system is governed by two reparametrization modes $t_L(u)$ and $t_R(u)$ whose effective action is given by a Schwarzian for each mode plus a contribution coming from the interaction term $H_\text{int}$ \eqref{eq:H_coupled}. In the zero-temperature limit, the solution for the saddle point is $t_{L/R}(u) = t'u$, from where one can derive the behavior of the two-point functions \cite{Maldacena:2018lmt}
\begin{align}
    \langle \chi^j_L(u)\chi^j_L(0)\rangle \sim e^{-i\pi\Delta}\left(\frac{t'}{\sin\frac{t'(u-i\epsilon)}{2}}\right)^{2\Delta}, \qquad \langle  \chi^j_L(u)\chi^j_R(0)\rangle \sim i \left(\frac{t'}{\cos\frac{t'u}{2}}\right)^{2\Delta},
\end{align}
where $\Delta = 1/q$ is the identified conformal dimension. Upon increasing the temperature, the system undergoes a phase transition interpreted as a transition from a wormhole to two weakly coupled black holes. At finite temperature, the system displays a similar oscillatory behavior in the two-point functions but there is also a decay in time \cite{Qi:2020ian}.  

We are interested in the behavior of the two-point functions for the sparse SYK at finite $N$. In this regime, the thermal average can be well approximated by an expectation value in a Haar-random state \cite{Goldstein:2005aib, Luitz:2017yps, Kobrin:2020xms}
\begin{equation} \label{eq:typicality}
    {\langle\chi_a^j(t)\chi_b^j(0)\rangle} =
    \frac{1}{Z}\text{Tr}\left[e^{-\beta H}\chi_a^j(t)\chi_b^j(0)\right]
    \simeq \frac{\langle\beta|\chi_a^j(t)\chi_b^j(0)|\beta\rangle}{\langle\beta|\beta\rangle},
\end{equation}
where $|\beta\rangle = e^{-\frac{\beta}{2}H}|\psi\rangle$ is the state obtained from imaginary time evolution of a random state $|\psi\rangle$ using the full coupled Hamiltonian $H$. This approximation follows from quantum typicality, and its error is expected to decrease exponentially with $N$ and temperatures above the gap of the system. Each component of $|\psi\rangle$ is drawn from a Gaussian distribution, and in practice the average is performed simultaneously over the random states and the disorder realizations of the couplings $J_{j_1,\ldots,j_q}^a$. Moreover, the all-to-all SYK is known to be self-averaging, meaning that when $N\to\infty$ a single disorder realization approaches the disorder average. For the sparse SYK system, however, the self-averaging property does not hold in general but it is still valid for the Green's function \cite{Xu:2020shn}.

\begin{figure}
    \centering
    \includegraphics[width=\linewidth]{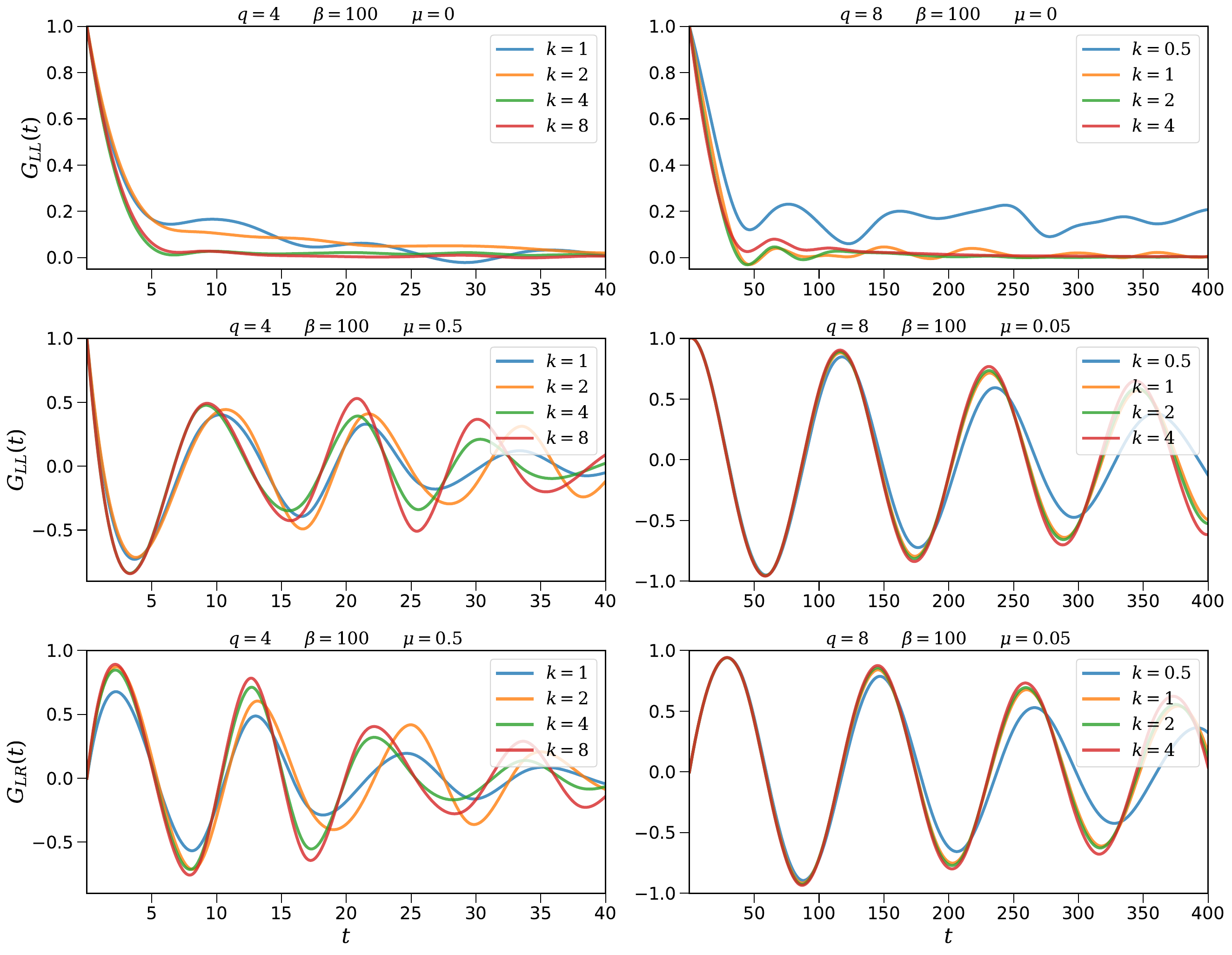}
    \caption{Real time retarded Green's function $G_{ab}(t)$ \eqref{eq:green-retarded} with $2N=40$, $q=4$ (Left) and $q=8$ (Right), for different choices of the sparsity parameter $k$. Each curve is an average over 50 disorder realizations of the random couplings. For $\mu=0$, the left-right correlator $G_{LR}$ (not shown) vanishes.}
    \label{fig:green-sparsity}
\end{figure}

The real time Green's function $G_{ab}(t)$ from our numerical simulation is shown in Fig.\,\ref{fig:green-sparsity}. In the uncoupled case at $\mu=0$, we obtain the typical decay of the SYK two-point function $G_{LL}\sim|t|^{-2/q}$. The addition of an interaction term makes the $G_{LR}$ correlator to be non-zero and produces an oscillatory behavior with the same characteristic frequencies in both $G_{LL}$ and $G_{LR}$. The existence of a non-zero left-right correlation indicates that signals can be transmitted between the left and right systems. This is not a surprise since we are explicitly adding an interaction between them, but what we are interested in is how these transmissions occur, and in particular whether or not they can be interpreted as a signal traveling through a traversable wormhole. We will explore these aspects in more detail in Section \ref{sec:transmission}.

Our main criteria in the estimation of the optimal sparsity is that it will give approximately the same characteristic frequency of oscillation for values of $k$ above the optimal sparsity. Empirically, we notice larger fluctuations for $q=4$ compared to the $q=8$ case as we vary the sparsity parameter $k$ within the same range. This suggests that the optimal sparsity for larger $q$ can be achieved using a smaller value of $k$, in agreement with our expectations based on the analysis of regular hypergraphs in Section \ref{sec:sparsity}. 

\begin{figure}
    \centering
    \includegraphics[width=\linewidth]{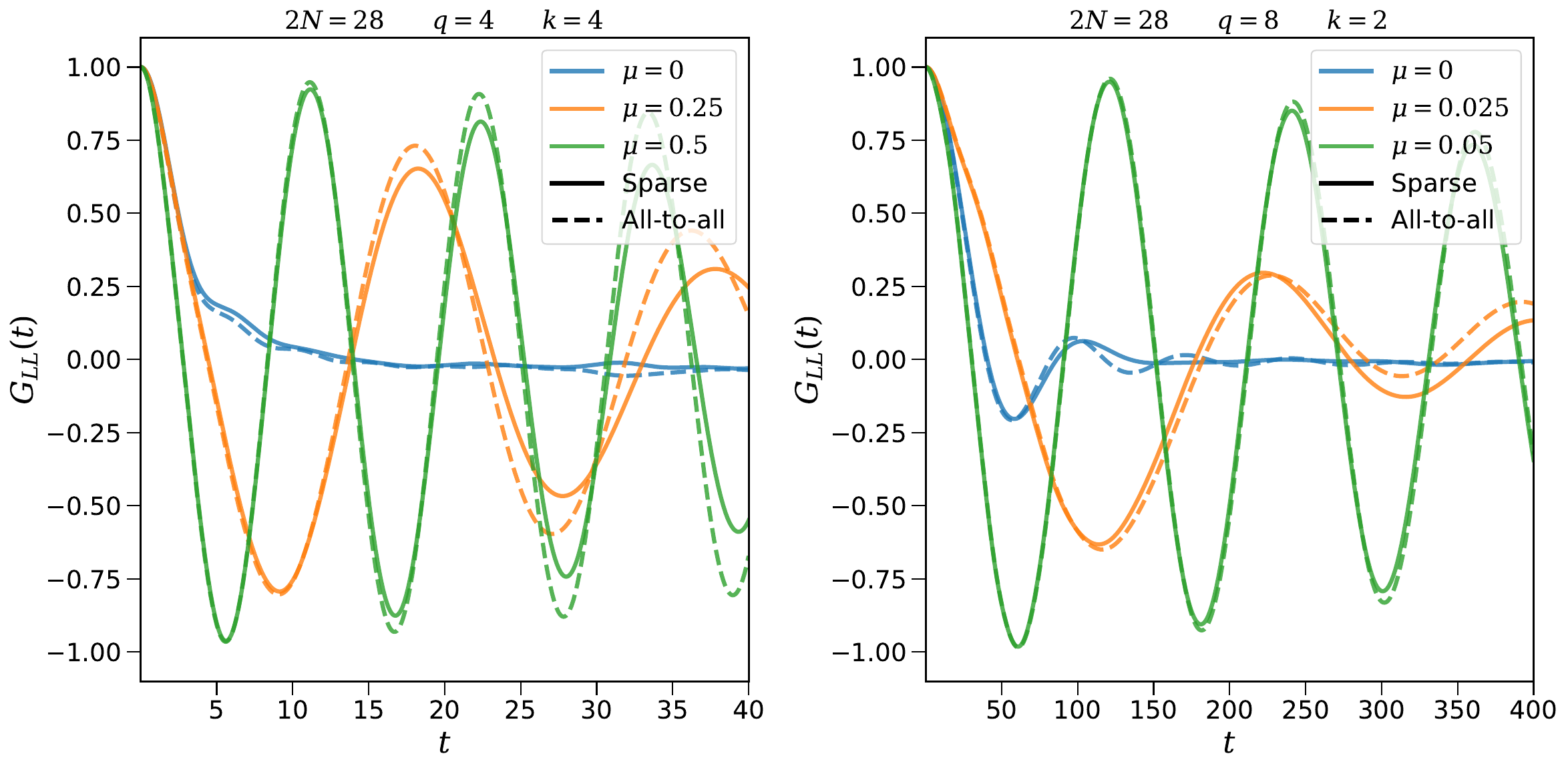}
    \caption{Retarded Green's function $G_{LL}$ for the two coupled sparse SYK model (solid) compared to the corresponding all-to-all SYK (dashed) for $2N=28$ for different values of the coupling $\mu$. We set $q=4$, $k=4$ (left) and $q=8$, $k=2$ (right), and each curve is obtained from an average over 100 realizations of the random couplings.}
    \label{fig:green-comp}
\end{figure}

For concreteness, we have selected the optimal sparsity to be $k=4$ for $q=4$, and $k=2$ for $q=8$. For these values, the behavior of the Green's function in the two coupled sparse SYK is in good agreement with the corresponding all-to-all model -- see Fig.\,\ref{fig:green-comp}. Even though the direct comparison with the all-to-all SYK in our simulation is restricted to small $N$, the hypergraph analysis in Section \ref{sec:sparsity} suggests that $k=\mathcal{O}(1)$ is enough to reproduce the all-to-all model regardless of the value of $N$. Thus, we expect that the selected optimal sparsity parameters will also be enough to recover the traversable wormhole features of the all-to-all coupled SYK at larger $N$. In what follows, we will support this choice by studying additional properties of the two coupled sparse SYK. 

\subsection{TFD and ground state overlap}

The thermofield double state (TFD) plays an important role in many constructions of holographic models. One natural question is how to actually build this state, in particular using the SYK model. In general, the Hamiltonian $H_L+H_R$ does not admit a TFD as its ground state. On the other hand, it has been shown that the TFD state can be approximately realized as the ground state of the two SYK system if we add the interaction $H_\text{int}$ \eqref{eq:coupledH} to the system \cite{Maldacena:2018lmt}. Thus, the TFD state for SYK can be prepared by starting from any excited state and then cooling the system down by coupling to an external bath \cite{Maldacena:2019ufo}. A variational approach whose goal is to minimize the expectation value of the coupled Hamiltonian has also been employed to simulate the TFD for the SYK model \cite{Su:2020zgc}. A more general discussion about how to prepare the thermofield double state can be found in Ref.\,\cite{Cottrell:2018ash}. 

While it has been understood that the ground state of the two coupled SYK system is approximately the TFD, it is not obvious that the same happens when we introduce sparsity in the SYK model. Here, we study in detail the properties of the ground state $|\Psi_0\rangle$ of the two coupled sparse SYK and its relation to the TFD state, by performing a similar analysis that was carried out for the all-to-all SYK \cite{Garcia-Garcia:2019poj, Lantagne-Hurtubise:2019svg, Sahoo:2020unu, Alet:2020ehp}. By taking advantage of the sparsity, we are able to reach up to $2N=48$ for $q=8$, which goes beyond previous studies for the corresponding all-to-all model \cite{Alet:2020ehp}.

The TFD state at inverse temperature $\beta$ is defined as
\begin{equation}
    |\text{TFD}_\beta\rangle = \sum_n e^{-\frac{\beta E_n}{2}}|E_n\rangle_L|E_n\rangle^*_R,
\end{equation}
where $|E_n\rangle_{L/R}$ are the eigenstates of the $L/R$ system and $*$ denotes complex conjugation. The TFD state is a purification of a thermal state, in the sense that the reduced density matrix $\rho_L = \text{Tr}_R|\text{TFD}_\beta\rangle\langle \text{TFD}_\beta|$ is a thermal density matrix of the left system and vice-versa.  The temperature of the approximated TFD obtained from the ground state of the coupled Hamiltonian \eqref{eq:coupledH} will depend on the parameter $\mu$ controlling the interaction between the two systems. When $\mu\to0$ we obtain the zero temperature TFD state, whereas $\mu\to\infty$ gives the infinite temperature TFD. 

We can determine the temperature of the TFD state by maximizing the overlap 
\begin{equation} \label{eq:overlap}
    |\langle \text{TFD}_{\beta(\mu)}|\Psi_0\rangle|,
\end{equation}
where $|\Psi_0\rangle$ is the ground state and we assume that both states are normalized. We follow closely the procedure adopted in Ref.\,\cite{Alet:2020ehp} which takes advantage of Krylov subspace techniques. First, we construct the coupled SYK Hamiltonian and find its ground state $|\Psi_0\rangle$. Then, we build the infinite temperature TFD state 
\begin{equation}
    |\text{TFD}_0\rangle \equiv |I\rangle.
\end{equation}
Note that $|I\rangle$ is just the ground state of $H_\text{int}$. The TFD at finite temperature can be obtained by imaginary time evolution using the uncoupled Hamiltonian
\begin{equation}
    |\text{TFD}_\beta\rangle \propto e^{-\delta\beta(H_L+H_R)/4}|I\rangle,
\end{equation}
with small increments $\delta\beta$ typically ranging from $0.01\mu\leq\delta\beta\leq 0.1\mu$. We keep evolving the system until the overlap $|\langle \text{TFD}_{\beta(\mu)}|\Psi_0\rangle|$ starts to decrease. Note that the imaginary time evolution can be carried out efficiently using Krylov subspace since we are not required to write the Hamiltonian matrix explicitly.

\begin{figure}
     \centering
     \begin{subfigure}[b]{0.49\textwidth}
         \centering
         \includegraphics[width=\textwidth]{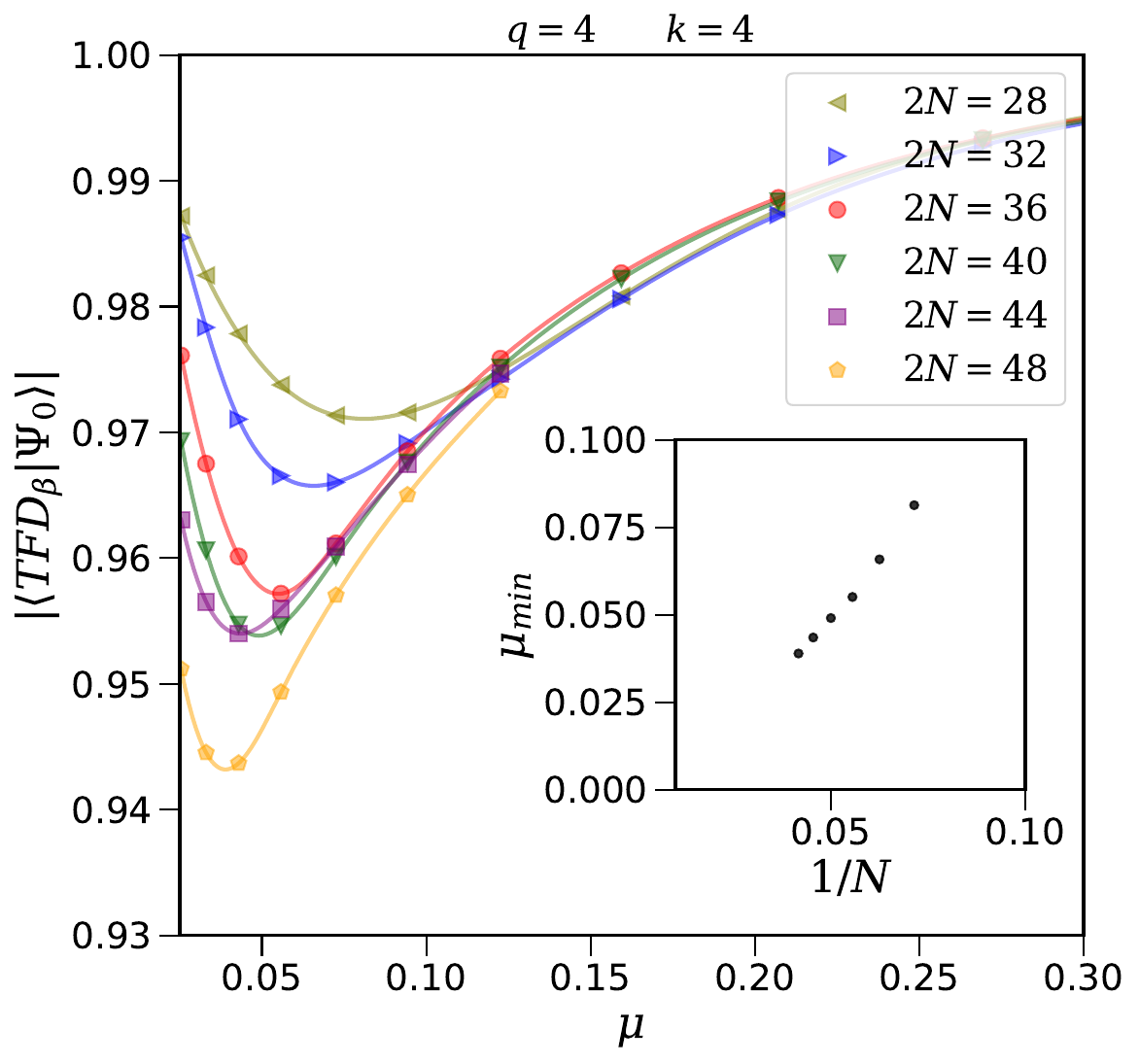}
     \end{subfigure}
     \hfill
     \begin{subfigure}[b]{0.49\textwidth}
         \centering
         \includegraphics[width=\textwidth]{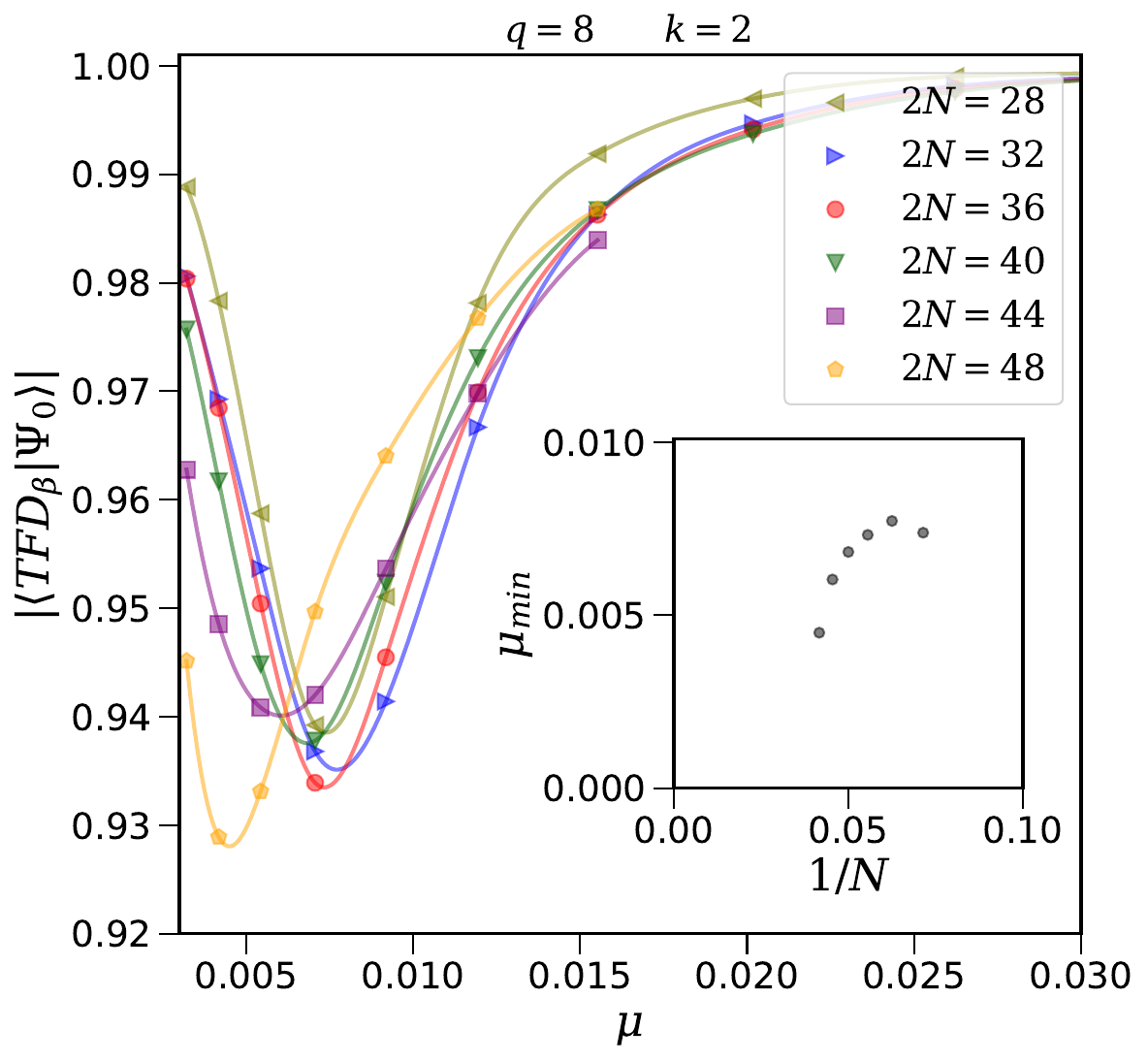}
    \end{subfigure}
     \caption{\textit{Main:} Maximum overlap between the ground state of two coupled sparse SYK model and the TFD, assuming that both states are normalized. \textit{Inset:} Finite $N$ scaling for the value $\mu_\text{min}$ at which the overlap reaches its minimum. We set $q=4$, $k=4$ (Left) and $q=8$, $k=2$ (Right), and performed an average over 50 disorder realizations.}
     \label{fig:overlap}
\end{figure}

In Fig.\,\ref{fig:overlap}, we show the (maximum) overlap obtained using the method described above. The overlap remains close to 1 for all values of the coupling strength, with a minimum $\gtrsim94\%$ for $q=4$ and $\gtrsim93\%$ for $q=8$. Our results are in good agreement with the results obtained for the all-to-all SYK where the overlap has been computed for up to $N=52$ Majoranas for $q=4$, and up to $N=40$ for $q=8$ \cite{Alet:2020ehp}. The sparsity in our model enabled us to push the simulation for $q=8$ to larger values of $N$ which suggests that $\mu_\text{min}$, the value of the coupling at the minimum overlap, goes to zero as $N\to\infty$, though more data would still be necessary to support this extrapolation.

\begin{figure}
     \centering
     \begin{subfigure}[b]{0.49\textwidth}
         \centering
         \includegraphics[width=\textwidth]{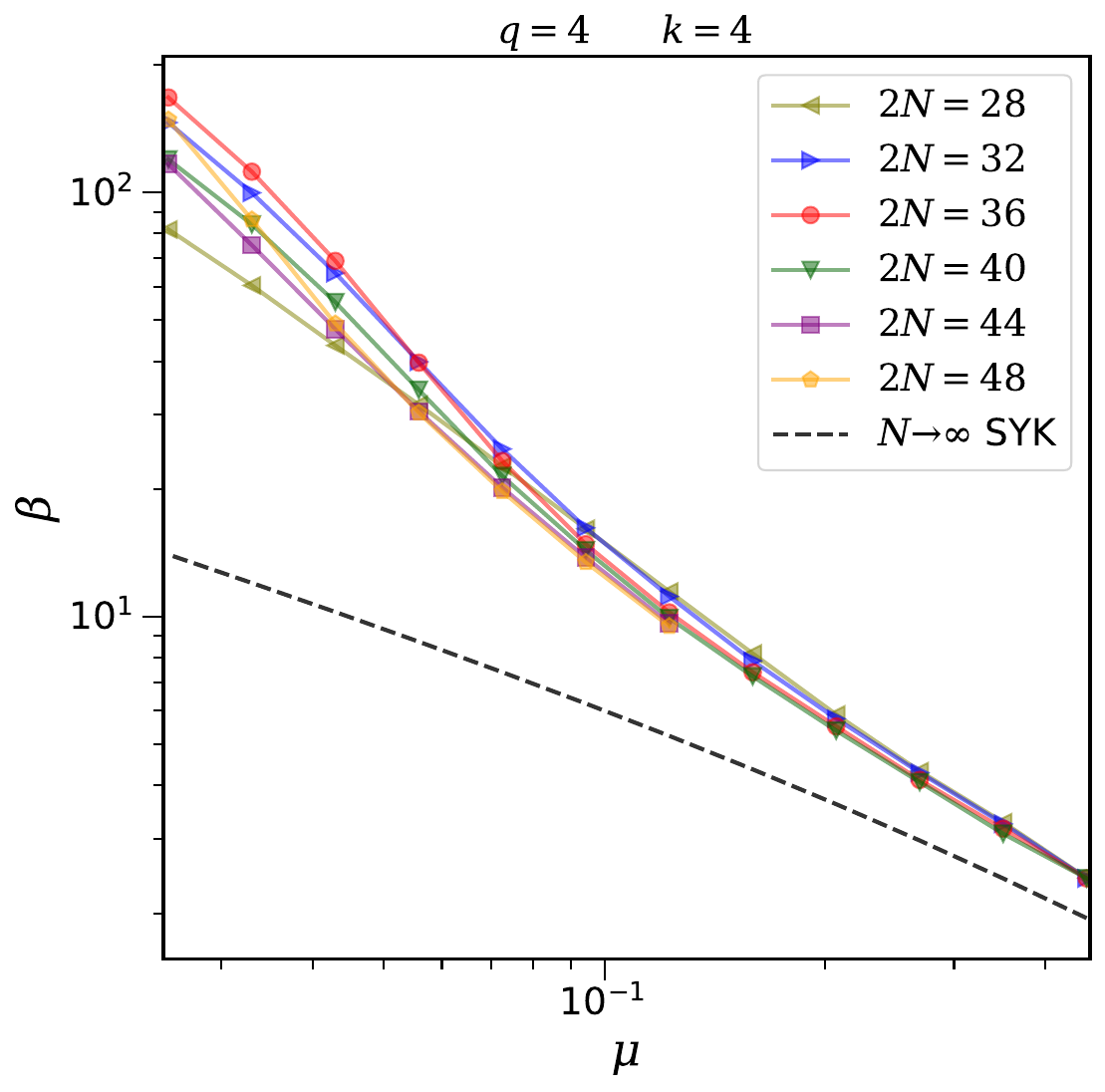}
     \end{subfigure}
     \hfill
     \begin{subfigure}[b]{0.49\textwidth}
         \centering
         \includegraphics[width=\textwidth]{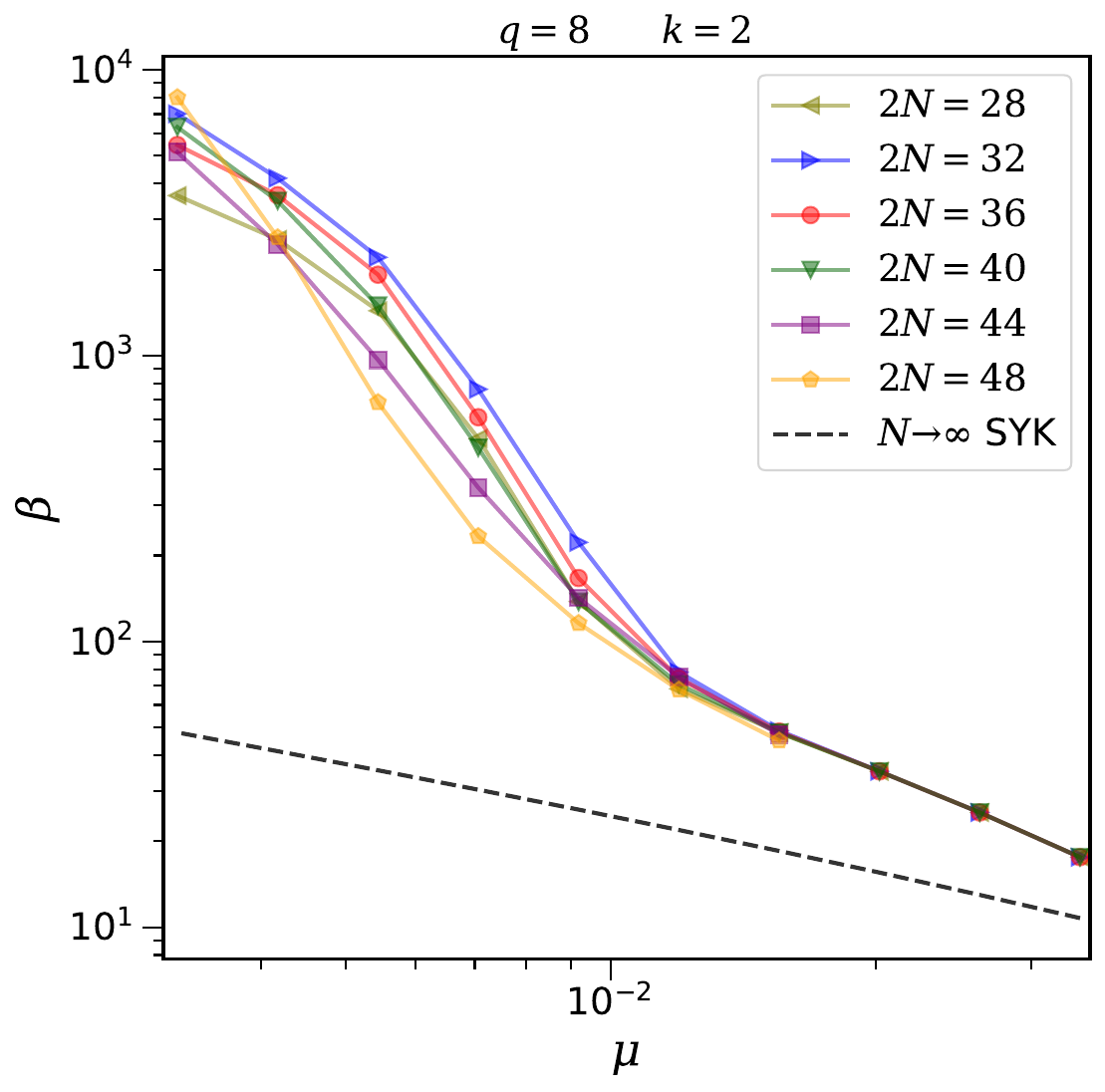}
     \end{subfigure}
     \caption{The inverse temperature $\beta(\mu)$ of the TFD state which maximizes the overlap $|\langle\text{TFD}_{\beta}|\Psi_0\rangle|$. We set $q=4$, $k=4$ (Left) and $q=8$, $k=2$ (Right), and performed an average over 50 disorder realizations. The dashed line corresponds to the analytical formula \eqref{eq:beta-mu} valid in the large $N$ and large $q$ limit.}
     \label{fig:beta}
\end{figure}

The temperature $\beta(\mu)$ of the TFD state that maximizes the overlap is presented in Fig.\,\ref{fig:beta}. We emphasize that the temperature of the TFD is simply an effective temperature and is not the same as the physical temperature of the system, which is zero since the system is in the ground state. In the large $N$ and large $q$ limit, it is possible to derive an analytical formula for $\beta(\mu)$, which is given by \cite{Maldacena:2018lmt} 
\begin{equation} \label{eq:beta-mu}
    \beta(\mu) = \frac{2}{\alpha} \sqrt{1+\left(\frac{\alpha}{\mathcal{J}}\right)^2}\arctan\frac{\mathcal{J}}{\alpha},
\end{equation}
where $ \alpha = \mathcal{J}\sinh\gamma$, and $\mu q = 2\alpha \tanh \gamma$. We found a large discrepancy when comparing our numerical results with the analytical formula \eqref{eq:beta-mu} in the small coupling region, signaling strong finite-size effects in this region. We start to obtain agreement with the analytical result for larger values of $\mu$. This behavior have already been observed for the all-to-all SYK \cite{Alet:2020ehp}, but in our simulation it becomes more evident that, for $q=8$, the numerical data gets closer to the analytical formula as we increase $N$, though the region of very small $\mu$ displays a non-monotonic behavior with respect to $N$ making it unclear how to draw a finite $N$ scaling in the small coupling region.

\subsection{Energy gap} \label{subsec:Egap}

The two coupled SYK model is known to be a gapped system, and its energy gap $E_\text{gap}$ provides a way to characterize different regimes of the system. In the traversable wormhole phase, the energy gap can be used to estimate the frequency of oscillations in the transmission amplitude of excitations between the two SYK systems \cite{Maldacena:2018lmt, Plugge:2020wgc}. The global $AdS_2$ picture \cite{Maldacena:2018lmt} suggests that in the low energy limit and at weak coupling, the energy gap is expected to scale as 
\begin{equation} \label{eq:Egap}
    E_\text{gap} \propto \mu^{\frac{1}{2-2\Delta}},
\end{equation}
where $\Delta=1/q$ is the conformal dimension. In particular we have $E_\text{gap} \propto \mu^{2/3}$ for $q=4$ and $E_\text{gap} \propto \mu^{4/7}$ for $q=8$. The appearance of this particular scaling of the energy gap at weak coupling and low energy regime can be obtained from the effective Schwarzian action that governs the dynamics of the system in this limit, from where one can derive that the spectrum contains a conformal tower of states
\begin{equation} \label{eq:Econf}
    E_n^\text{conf} = t'(\Delta+n), 
\end{equation}
where $t'\sim \mu^\frac{1}{2-2\Delta}$. This is the same as the spectrum for a bulk field in $AdS_2$. There is also an extra quantum mechanical degree of freedom referred to as `boundary graviton', which encodes the gravitational dynamics and backreaction of the system. Its spectrum sector is given by
\begin{equation}
    E_n^\text{bg} = t'\sqrt{2(1-\Delta)}\left(n+\frac{1}{2}\right),
\end{equation}
Thus, it follows that the system has an energy gap scaling as \eqref{eq:Egap}. On the other hand, at strong coupling the interaction term in \eqref{eq:coupledH} dominates and we expect the linear scaling
\begin{equation} \label{eq:Egap_strong}
    E_\text{gap}\propto \mu.
\end{equation}
simply because in this regime the system behaves roughly as a free fermionic harmonic oscillator.

The energy gap $E_\text{gap}$ provides a way to identify distinctive features of the traversable wormhole phase. In the large $N$ limit, the energy gap can be extracted from the decay rate of the imaginary time Green's function \cite{Maldacena:2018lmt}. In the wormhole phase, the decay of the Left-Left and Left-Right Green's function is exponential and their decay rates are equal and given precisely by the energy gap. 

\begin{figure}
     \centering
     \begin{subfigure}[b]{0.49\textwidth}
         \centering
         \includegraphics[width=\textwidth]{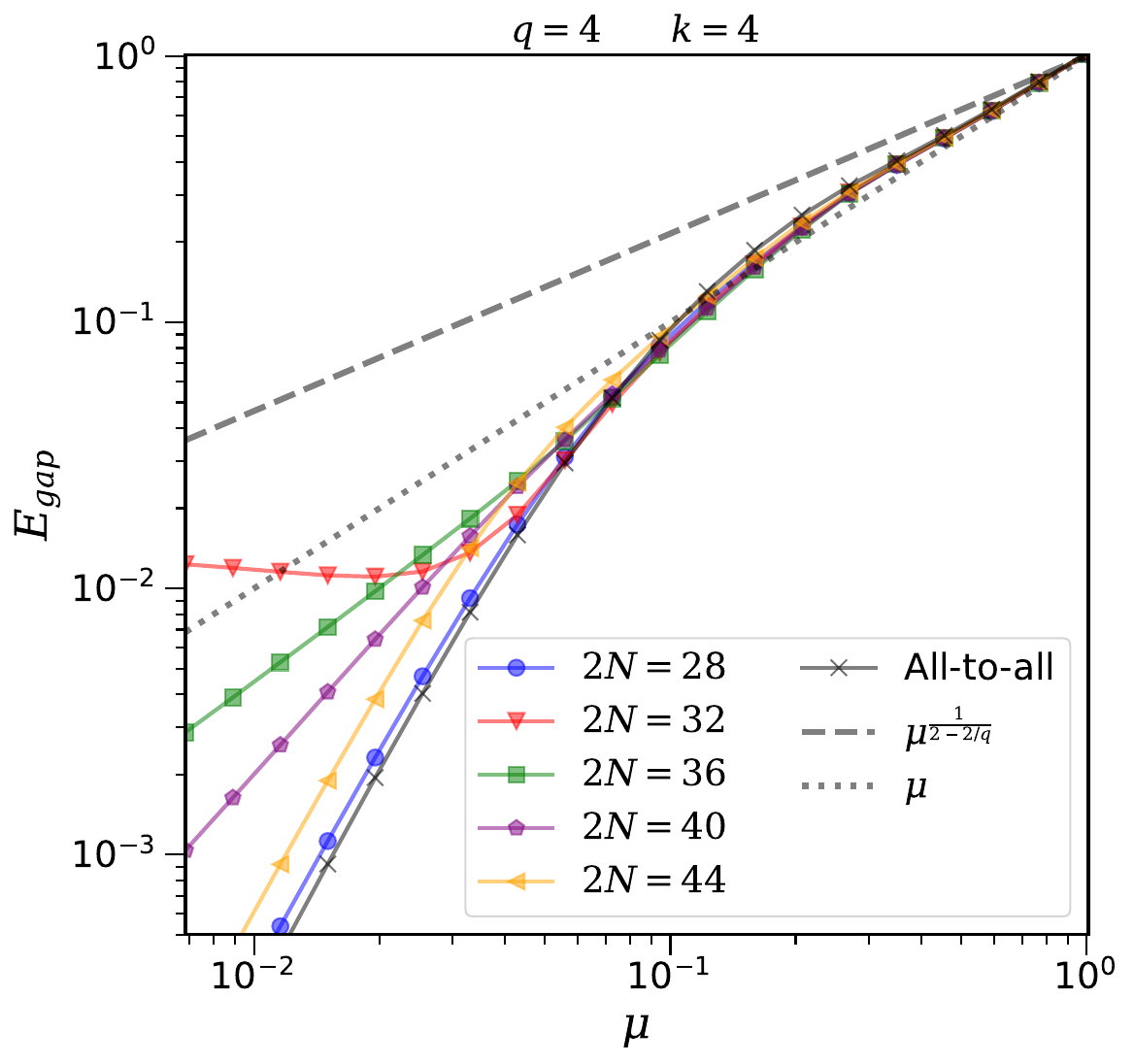}
     \end{subfigure}
     \hfill
     \begin{subfigure}[b]{0.49\textwidth}
         \centering
         \includegraphics[width=\textwidth]{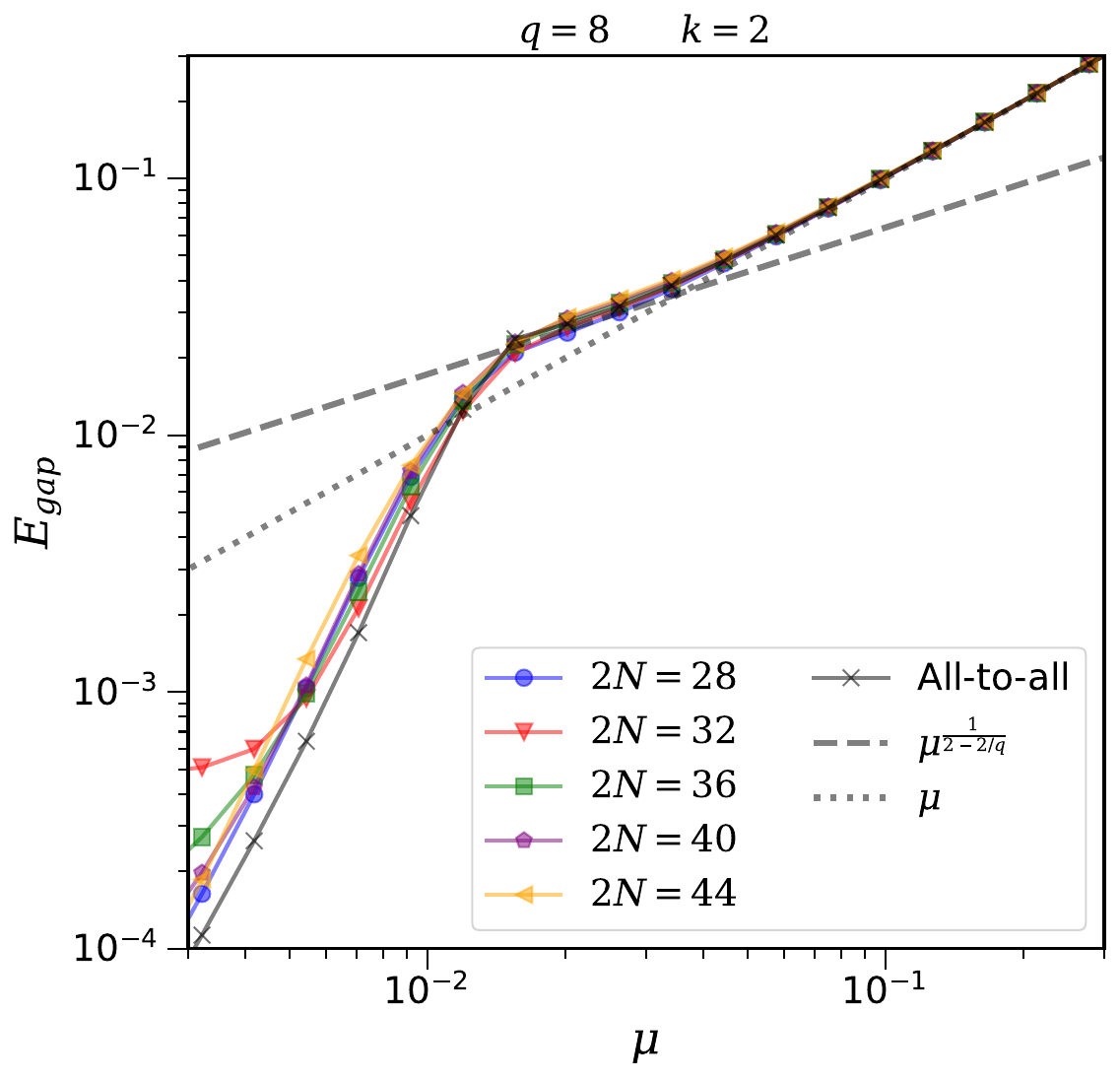}
     \end{subfigure}
     \caption{Energy gap of the two coupled sparse SYK system for $q=4$, $k=4$ (Left) and $q=8$, $k=2$ (Right). Dotted and dashed lines are the expected scalings $E_\text{gap}\propto\mu^\frac{1}{2-2/q}$ and $E_\text{gap}\propto\mu$ at weak and strong coupling, respectively, obtained from the large $N$ analysis. The energy gap for the two coupled all-to-all SYK model with $2N=28$ is also included for comparison. The results correspond to an average over 50 disorder realizations.}
     \label{fig:gap}
\end{figure}

At finite $N$, though, we can extract the energy gap from direct computation of the lowest energy levels of the two coupled SYK system. In Fig.\,\ref{fig:gap} we show our results for the energy gap obtained in that way. For $q=4$, we find agreement with results using exact diagonalization for the all-to-all SYK \cite{Lantagne-Hurtubise:2019svg}. At large enough couplings we obtain the expected linear scaling. At very small couplings, however, we observe large variations with respect to $N$, signaling that finite $N$ effects are dominant in that region. 

Remarkably, at some intermediate coupling, we can match the scaling predicted from holography \eqref{eq:Egap}. This is more visible for $q=8$, where this scaling is obeyed within the range $0.015 \lesssim \mu \lesssim 0.05$. In the next section we will study the transmission of signals between the two sparse SYKs and show the transmission can be related to the energy gap.

\section{Diagnostics of signal transmission}
\label{sec:transmission}

The Maldacena-Qi model \cite{Maldacena:2018lmt} has a wormhole phase in the regime of small $\mu/J$ and low-energy. In the traversable wormhole picture, it is straightforward to interpret the transmission of a signal from one boundary to the other simply as the signal traversing the wormhole. On the other hand, from the quantum mechanical point of view, this phenomena can be understood as a `revival', in which a perturbation in one side of the system gets scrambled and after some time, due to the effect of the interaction, it reassembles itself on the other system \cite{Gao:2018yzk}. 

In the two coupled SYK model, the revival phenomena has been investigated for $q=4$ at both small and large $N$ \cite{Plugge:2020wgc}. In the large $N$ limit, revival oscillations at frequency $\omega \propto \mu^{2/3}$ have been observed by solving the Schwinger-Dyson equations, in agreement with predictions from holography. On the other hand, for small $N$, an emergent conformal tower of states \eqref{eq:Econf} could not be observed and revival oscillations arise from a different mechanism attribute to finite-size effects. It remained unclear how the dynamics change with increasing $N$ giving rise to a traversable wormhole behavior. 

Here, we make use of Krylov subspace methods combined with the introduction of sparsity in the Hamiltonian to push this investigation to larger values of $N$. We study the sparse two coupled SYK for $2N=40$, for $q=4,8$ with sparsity parameter $k=4,2$, respectively, at both zero and finite temperature. We again consider the two coupled sparse SYK Hamiltonian \eqref{eq:H_coupled}
\begin{equation}
    H = H_{L} + H_{R} + i\mu\sum_j \chi_L^j \chi_R^j,
\end{equation}
where $H_{a}$ is the sparse SYK Hamiltonian with $N$ Majorana fermions for the $a=L,R$ system. We will be interested in the real time greater Green's functions
\begin{equation}
    G^>_{ab}(t) = -\frac{i}{N}\sum_j\langle\mathcal{T} \chi_a^j(t)\chi_b^j(0)\rangle =
    \begin{pmatrix}
        G^>_{LL}(t) & G^>_{LR}(t) \\
        G^>_{RL}(t) & G^>_{RR}(t) 
    \end{pmatrix}.
\end{equation}
For Majorana fermions, the lesser and greater Green's functions are related by $ G_{ab}^<(t)=(G_{ab}^>(t))^*$, and due to time translation symmetry we have $G_{LL}^>(t)=G_{RR}^>(t)$ and $G_{LR}^>(t)=-G_{RL}^>(t)$. From the greater Green's function we can obtain the transmission amplitude as 
\begin{equation} \label{eq:Tab}
    T_{ab} = 2|G^>_{ab}|.
\end{equation}
The square of the transmission amplitude reflects the probability of recovering $\chi_a^j$ at some time $t$ after inserting $\chi_b^j$ at $t=0$. Another quantity of interest is the spectral function $\rho_{ab}(\omega)$, which is related to ${G}_{ab}^>(t)$ via
\begin{equation} \label{eq:rho}
    \tilde{G}_{ab}^>(\omega) = (1-n_F(\omega))\rho_{ab}(\omega),
\end{equation}
where $\tilde{G}_{ab}^>(\omega)$ is the Fourier transform of ${G}_{ab}^>(t)$ and $n_{F}(\omega)=1/(1+e^{\beta\omega})$ is the Fermi-Dirac function.

\subsection{Transmission at zero temperature} \label{subsec:zero}

We first consider the transmission amplitude in the zero temperature limit $\beta\to\infty$, i.e., the transmission amplitude from excitations of the ground state. We can imagine that, at time $t=0$, we create a Majorana excitation on the Right system so that the full state of the system is
\begin{equation} \label{eq:excitation}
    |\Psi(t=0)\rangle = \chi_R|\Psi_0\rangle \equiv |\Psi_R\rangle.
\end{equation}
As time evolves, the excitation gets scrambled. According to the revival phenomena \cite{Gao:2018yzk, Plugge:2020wgc}, after some time $t_\text{rev}$, the excitation can reassemble itself and become localized on the Left system so that the quantum state will take the form
\begin{equation}
    |\Psi(t=t_\text{rev})\rangle = \chi_L|\Psi_0\rangle \equiv |\Psi_L\rangle.
\end{equation}
In principle, this process repeats itself with Left and Right interchanged and one can imagine it as an excitation that travels back and forth between the Left and Right systems. This is what we refer to as revival oscillations. The transmission amplitudes are then given by
\begin{equation} \label{eq:Tzero}
    T_{LR}(t) = |\langle\Psi(t)|\Psi_R\rangle|, \quad T_{LL}(t) = |\langle\Psi(t)|\Psi_L\rangle|.
\end{equation}

\begin{figure}
    \centering
    \includegraphics[width=\linewidth]{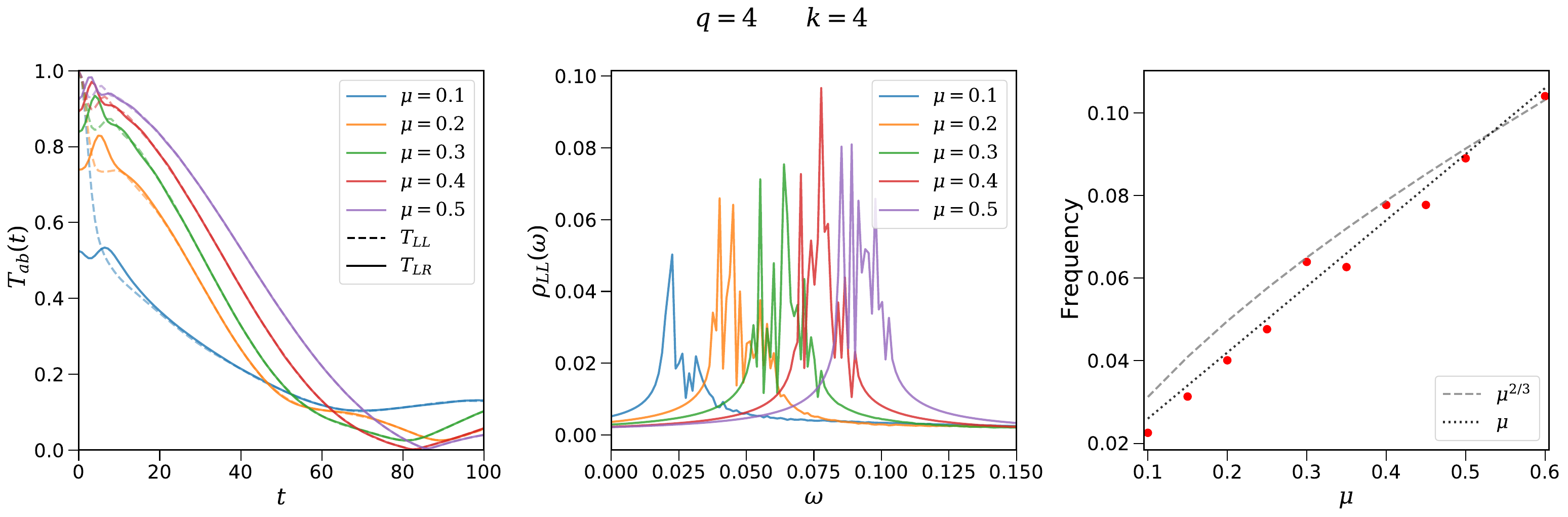}
    \includegraphics[width=\linewidth]{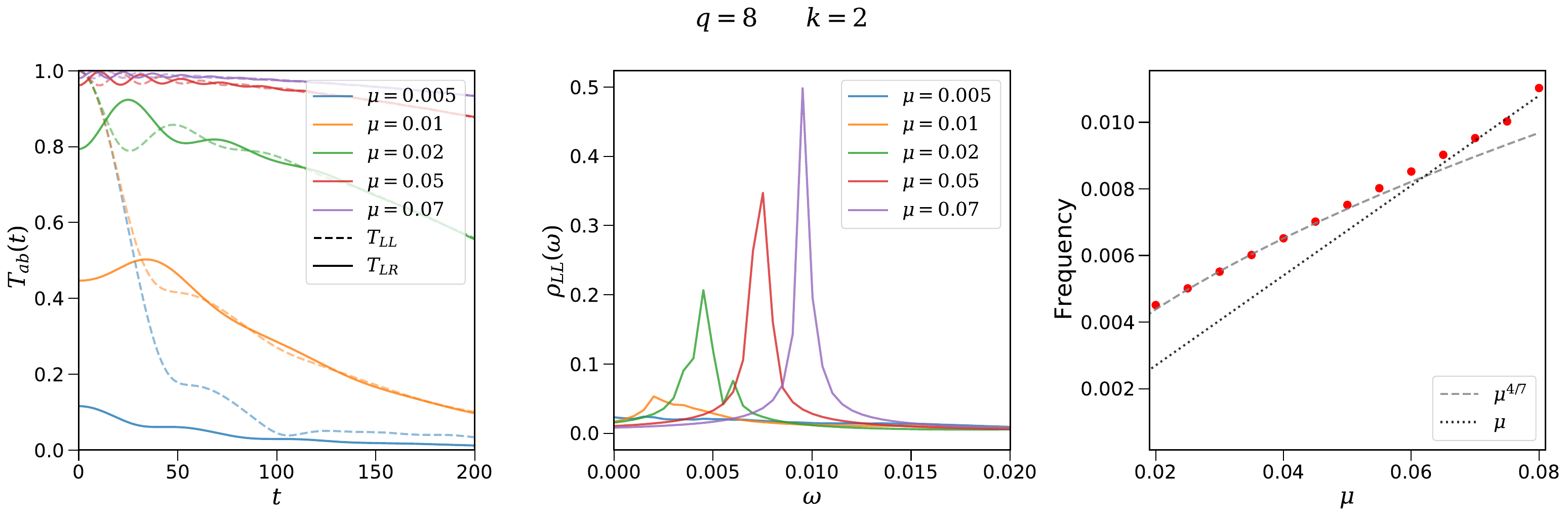}
    \caption{Transmission amplitude $T_{ab}$ at zero temperature \eqref{eq:Tzero} (left), the spectral function $\rho_{LL}$ \eqref{eq:rho} (center), and the characteristic frequency of oscillations of the Green's function (right) for $2N=40$, $q=4$, $k=4$ (top) and $q=8$, $k=2$ (bottom). The dashed and dotted lines represent the expected frequency scaling at weak and strong coupling, respectively, according to \eqref{eq:Egap} and \eqref{eq:Egap_strong}. We performed an average over 50 disorder realizations.}
    \label{fig:Gzero}
\end{figure}

In Fig.\,\ref{fig:Gzero}, we show the transmission amplitude at zero temperature for $q=4$ and $q=8$ for the two coupled system with $2N=40$. The expected revival oscillation, in which the transmission amplitudes $T_{LR}$ and $T_{LL}$ alternate their local minimum and maximum, is not observed for $q=4$ but can be noticed for $q=8$ at early times. This is presumably due to the suppression in the transmission amplitude from the typical decay $|t|^{-2/q}$ of the Green's function. 

Remarkably, for $q=8$, the spectral function $\rho_{LL}$ \eqref{eq:rho} displays a sharp peak whose frequency obeys the expected scaling $\propto \mu^\frac{1}{2-2/q}$ for the range of couplings $0.02 \lesssim \mu \lesssim 0.06$, and it changes to a linear scaling for $\mu\gtrsim0.06$. On the other hand, the position of the highest peak in the spectral function for $q=4$ is closer to a linear scaling, although the appearance of multiple frequencies in the spectral function prevent us to obtain a precise scaling. Moreover, these multiple frequencies are not separated enough to be compatible with the spectrum \eqref{eq:Econf} predicted by holography.

\begin{figure}
    \centering
    \begin{subfigure}[b]{0.32\textwidth}
        \centering
        \includegraphics[width=\textwidth]{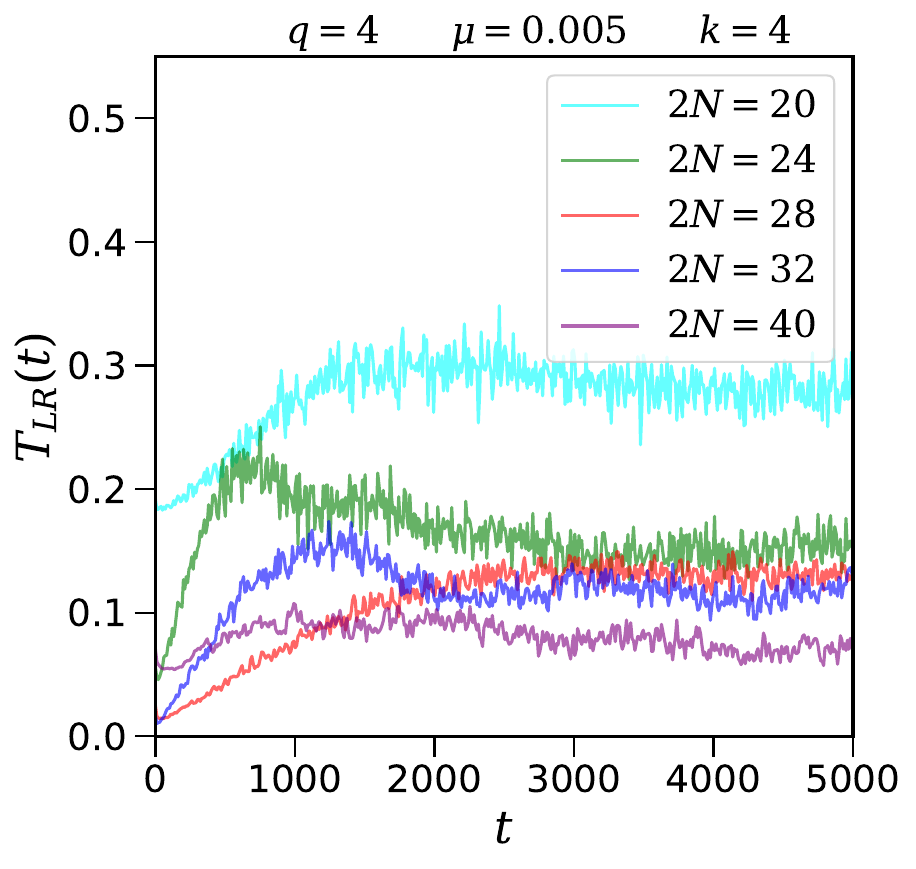}
    \end{subfigure}
    \hfill
    \begin{subfigure}[b]{0.32\textwidth}
        \centering
        \includegraphics[width=\textwidth]{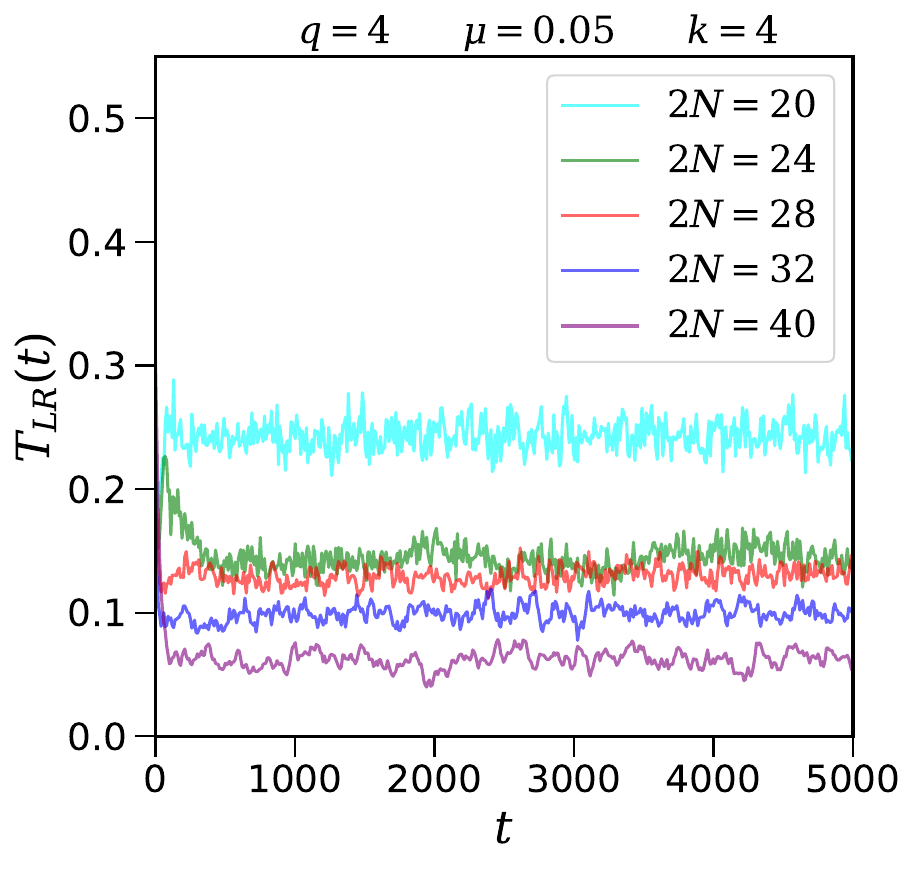}
    \end{subfigure}
    \hfill
    \begin{subfigure}[b]{0.32\textwidth}
        \centering
        \includegraphics[width=\linewidth]{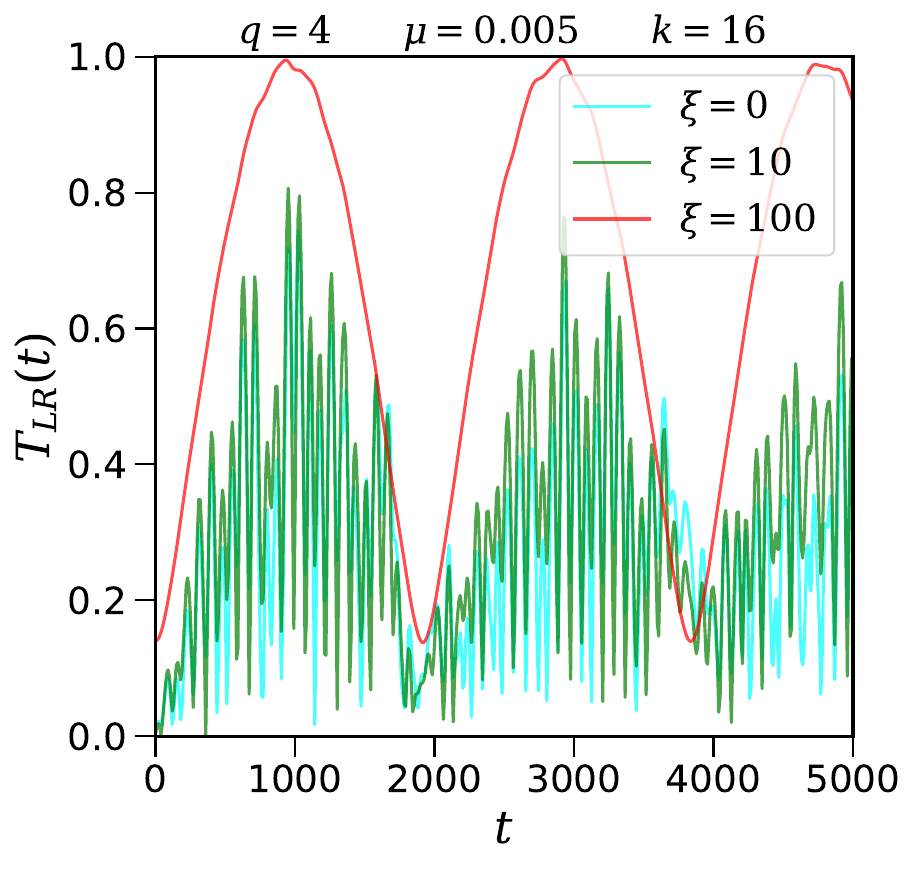}
    \end{subfigure}
    \caption{Transmission amplitude $T_{LR}$ at zero temperature for fixed sparsity parameter $k=4$ and different values of $N$, averaged over 50 disorder realizations, for (a) $\mu=0.005$ and (b) $\mu=0.05$. (c) Single realization of the transmission amplitude $T_{LR}$ for low-energy states with $2N=32$ and sparsity parameter $k=16$.}
    \label{fig:transmission-q4}
\end{figure}

Below some value of the coupling, which is approximately $\mu\lesssim 0.15$ for $q=4$ and $\mu\lesssim 0.015$ for $q=8$, the oscillatory behavior in the Green's function is no longer observed. This value is around the point at which the overlap between the TFD and the ground state \eqref{eq:overlap} starts to decrease (as we decrease $\mu$), which is also near the point where the energy gap starts to deviate from the large $N$ expectation. In this region of smaller couplings, a different type of signal transmission that is not connected to the wormhole phase can be observed \cite{Plugge:2020wgc}. In Fig.\,\ref{fig:transmission-q4}, we show that a non-zero transmission amplitude $T_{LR}$ is also observed at late times, which also indicates that the system does not thermalize. We found agreement with a numerical simulation carried out using exact diagonalization for the all-to-all SYK \cite{Plugge:2020wgc}, and our results for larger $N$ also supports the conclusion that this type of transmission is a particular feature of the finite $N$ regime, as the transmission amplitude clearly decreases with $N$. 

Also in the small coupling regime described above, revival oscillations can be obtained by considering low-energy excitations, i.e., instead of looking at a single Majorana excitation \eqref{eq:excitation} we consider its projection into low-energy states
\begin{equation}
    |\Psi_R\rangle_\text{low-energy} = \frac{1}{\sqrt{C}}\sum_{n=0}^{n_\text{max}}e^{-\xi\frac{(E_n-E_0)}{|E_0|}}\langle \Psi_n|\chi^j_R|\Psi_0\rangle |\Psi_n\rangle,
\end{equation}
where $|\Psi_n\rangle$ are the low-energy eigenstates of the coupled system, $\xi>0$ is a parameter controlling the suppression of higher energy states, and $C$ is a normalization constant. We have also introduced a parameter $n_\text{max}\lesssim 100$ because in practice we cannot solve for the whole spectrum when $N$ is large enough. Fig.\,\ref{fig:transmission-q4}c shows an example of signal transmission for a low-energy state for $2N=32$ and $q=4$. Since this type of transmission involves a larger time range compared to our previous analysis, quantitative agreement is obtained only for $k=16$. As a side remark, this type of transmission amplitude has a strong dependence on the value of $N$ and could possibly be related to its random matrix theory (RMT) classification.\footnote{Depending on the value of $N \text{ mod } 8$, the SYK model is classified as a Gaussian Orthogonal Ensemble (GOE), Gaussian Unitary Ensemble (GUE), or Gaussian Symplectic Ensemble (GSE) \cite{You:2016ldz}.} It would be interesting to understand if there is a connection between this type of revivals and the RMT universality classes.

\subsection{Finite temperature effects}

We now turn our attention to the transmission amplitude at finite temperature. The two coupled SYK model with a given fixed coupling $\mu$ undergoes a first order phase transition as the temperature of the system increases. This has been interpreted as a transition from a wormhole phase to a black hole phase \cite{Maldacena:2018lmt}. Intuitively, at higher temperatures the effect of the interaction $H_\text{int}$ \eqref{eq:coupledH} becomes less important, in the renormalization group sense, leading to two high temperature, weakly coupled black holes. At larger couplings, the phase transition takes place at higher temperatures. At some coupling $\mu_\text{max}$, the phase transition terminates. 

Several aspects of the two coupled SYK model at finite temperature have been investigated by solving the Schwinger-Dyson equations at both small and large $q$ in Ref. \cite{Qi:2020ian}. Our goal here is to study the temperature dependence of the Green's functions and how they affect the transmission amplitude at finite $N$. In particular, we want to analyze whether or not a phase transition occurs at finite $N$ in our sparse two coupled SYK model. 

\begin{figure}
    \centering
    \begin{subfigure}[b]{0.32\textwidth}
        \centering
        \includegraphics[width=\textwidth]{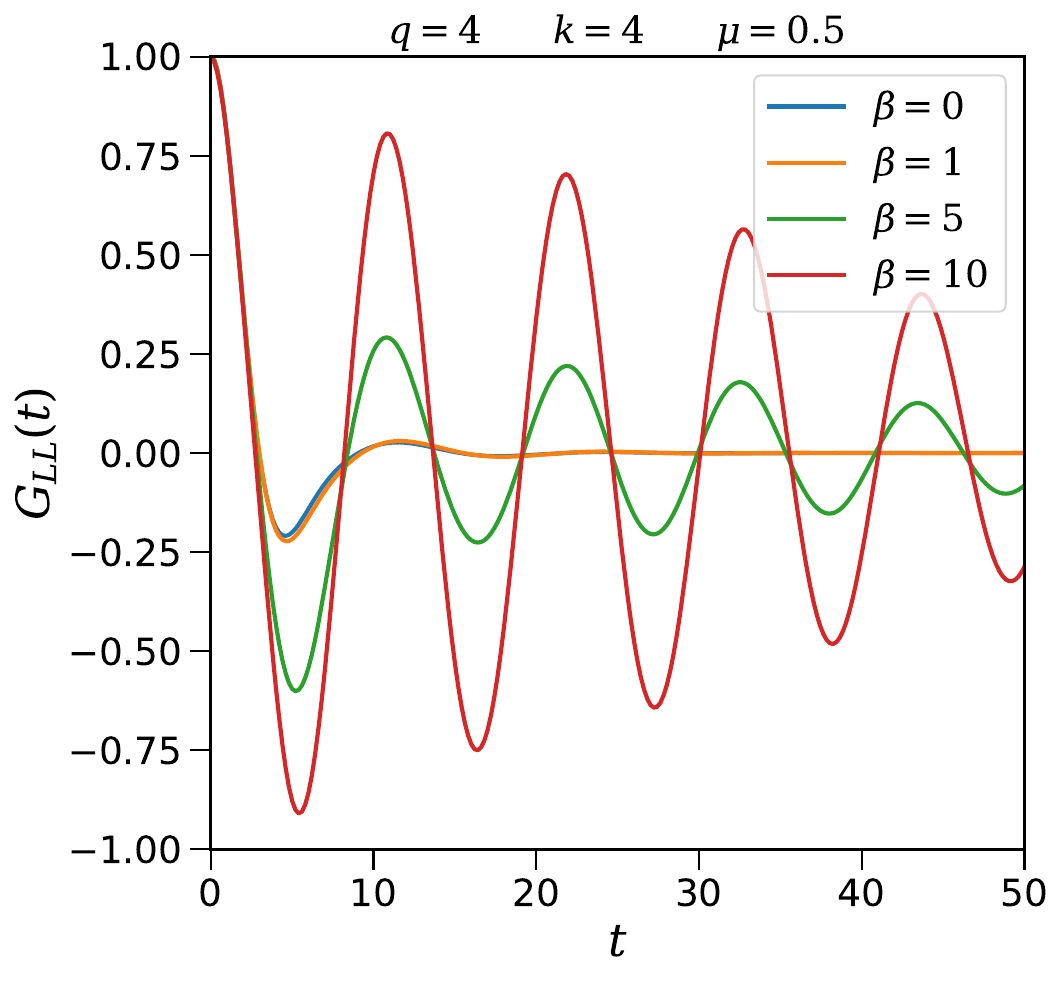}
    \end{subfigure}
    \hfill
    \begin{subfigure}[b]{0.32\textwidth}
        \centering
        \includegraphics[width=\textwidth]{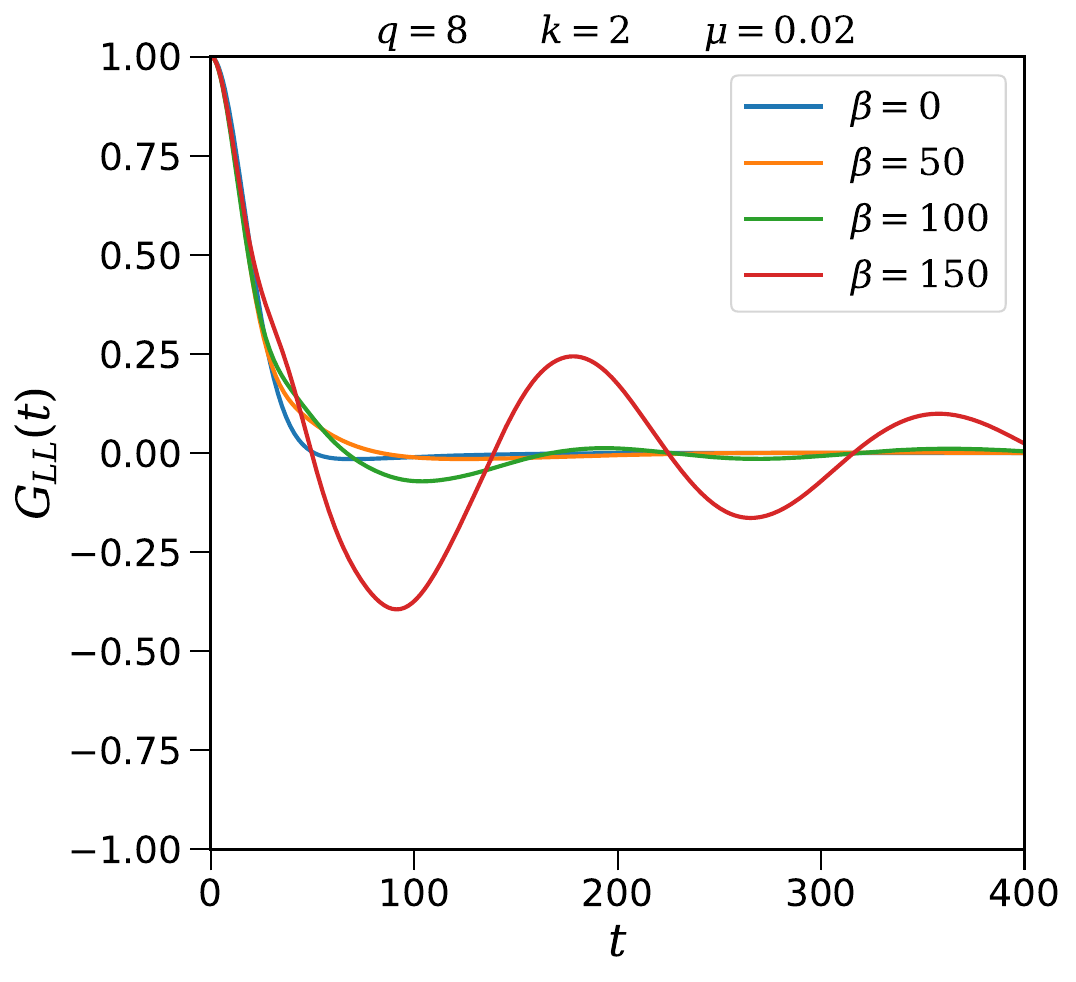}
    \end{subfigure}
    \hfill
    \begin{subfigure}[b]{0.32\textwidth}
        \centering
        \includegraphics[width=\linewidth]{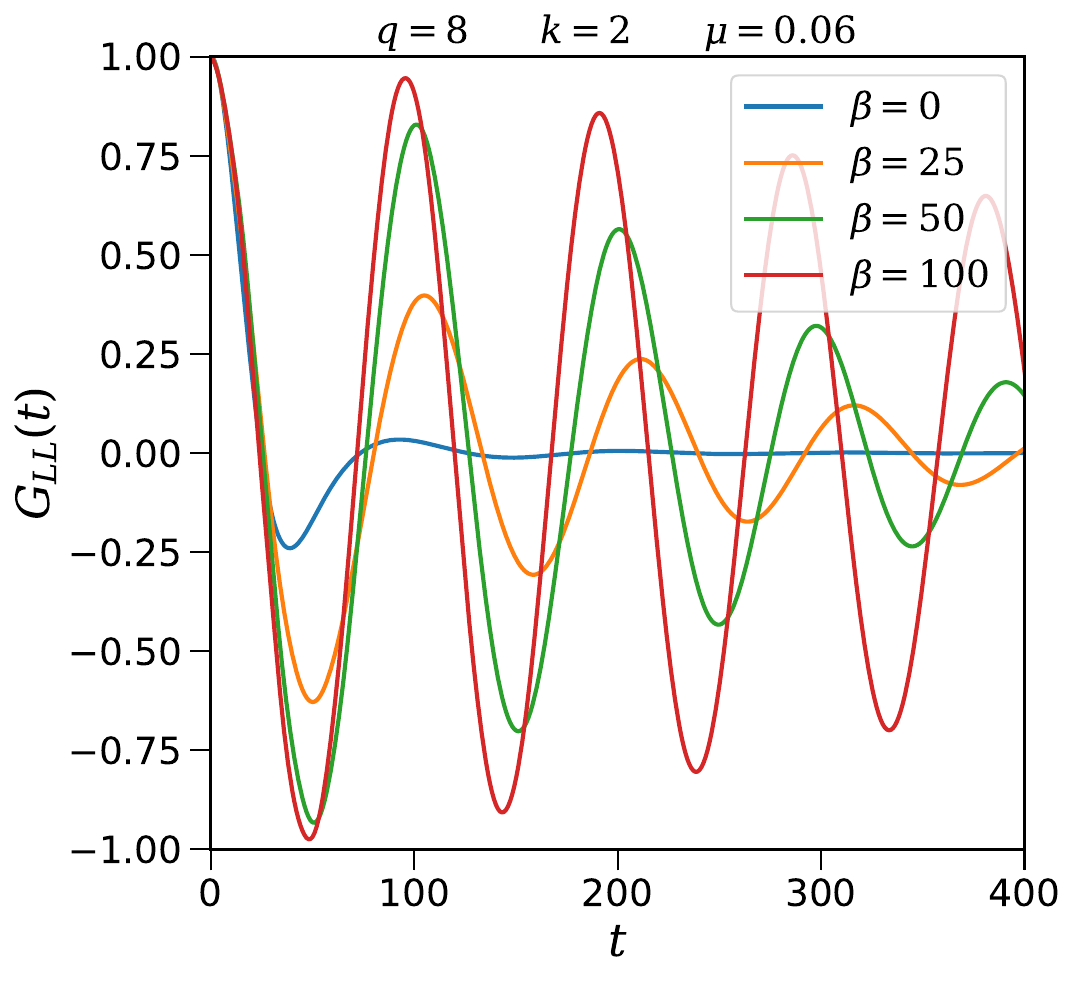}
    \end{subfigure}
    \begin{subfigure}[b]{0.32\textwidth}
        \centering
        \includegraphics[width=\textwidth]{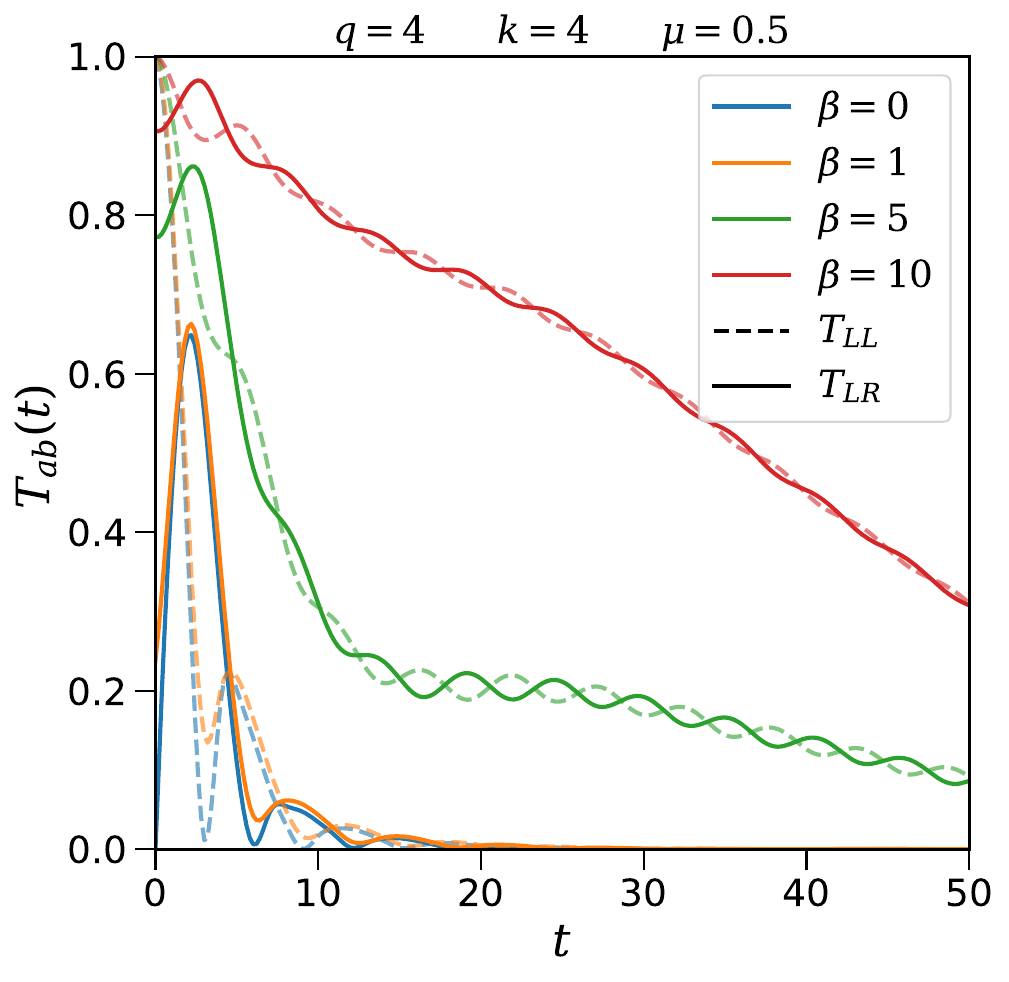}
    \end{subfigure}
    \hfill
    \begin{subfigure}[b]{0.32\textwidth}
        \centering
        \includegraphics[width=\textwidth]{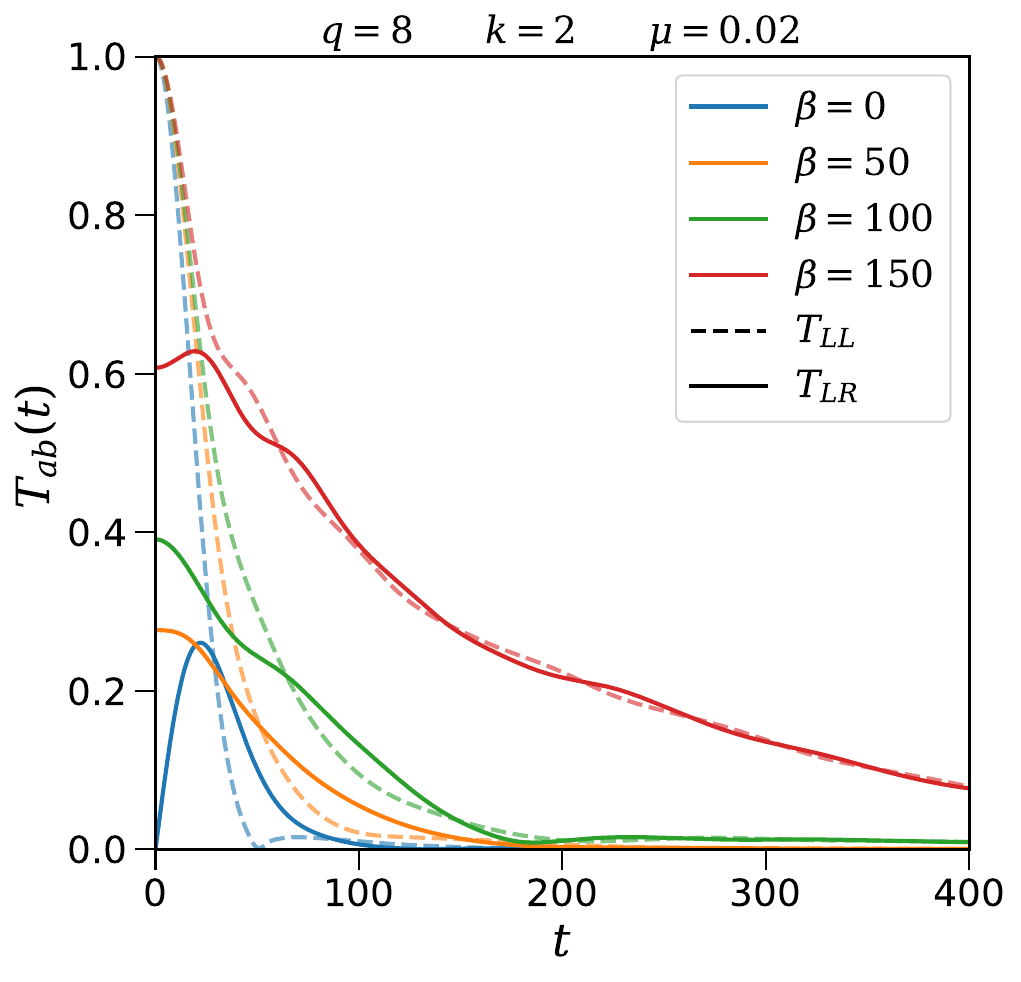}
    \end{subfigure}
    \hfill
    \begin{subfigure}[b]{0.32\textwidth}
        \centering
        \includegraphics[width=\linewidth]{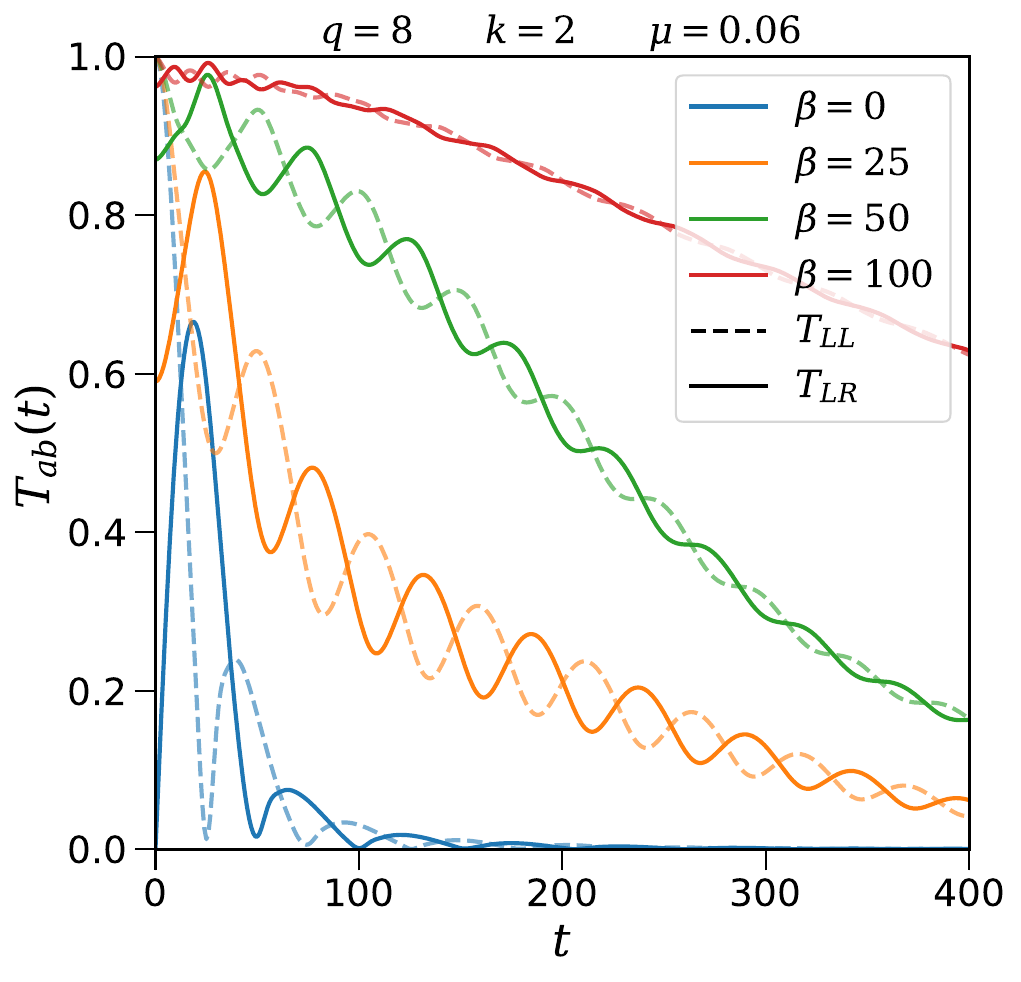}
    \end{subfigure}
    \caption{The effect of temperature on real time retarded Green's function $G_{LL}$ (top) and the corresponding transmission amplitude $T_{ab}$ (bottom) for the two coupled sparse SYK system with $2N=40$ Majorana fermions and parameters: $q=4$, $k=4$, and $\mu=0.5$ (left);  $q=4$, $k=4$, and $\mu=0.5$ (center); $q=4$, $k=4$, and $\mu=0.5$ (right). Each curve is the result of an average over 50 disorder realizations.}
    \label{fig:Tbeta}
\end{figure}

We numerically evaluated the retarded Green's function $G_{LL}$ \eqref{eq:green-retarded} and the transmission amplitude $T_{ab}$ \eqref{eq:Tab} as a function of the temperature using Krylov subspace techniques combined with the approximation of the thermal average by a pure state \eqref{eq:typicality}. Our results for $2N=40$ are shown in Fig.\,\ref{fig:Tbeta}. We first notice that, unlike the zero temperature case, the system thermalizes since $T_{ab}\to0$ for sufficiently large times. In general, as we increase the temperature of the system, the Green's function changes qualitatively from a non-oscillating function with the typical SYK decay to an oscillating function whose amplitude also increases with the temperature. This is similar to the transition between the wormhole and black hole phases discussed above. Although this is a good evidence that the phase transition is also present at finite $N$, a more detailed study of the thermodynamic properties is required to fully characterize the phase diagram of the sparse two coupled system.

The signature of revival oscillations, with $T_{LL/LR}$ alternating their local minimum and maximum, is enhanced at higher temperatures. Revival oscillations can be observed also at infinite temperature for a short time interval, as it can be seem in our result for $q=8$ and $\mu=0.06$ in Fig.\,\ref{fig:Tbeta}. This particular value of the coupling is approximately the value at which we observed a change of from $\propto \mu^\frac{1}{2-2/q}$ to linear scaling in the zero-temperature analysis in Section \ref{subsec:zero}. This indicates that a phase transition is no longer present, in agreement with previous studies \cite{Maldacena:2018lmt, Garcia-Garcia:2019poj}.

Transmission in the infinite temperature in the SYK model have already been observed in a recent quantum teleportation protocol \cite{Gao:2019nyj}. It has been found that varying the temperature of the SYK model gives a continuous interpolation between teleportation at low temperature and the so called peaked-size teleportation at high temperature \cite{Schuster:2021uvg}.  It would be interesting to investigate if there is a direct connection between these protocols and the transmission that we observed here.

\section{Discussion}
\label{sec:discussion}

In this work we investigated a variant of the SYK model defined on a sparse regular hypergraph. In this model, the Hamiltonian is a sum of exactly $k\,N$ terms, in contrast to the $\sim N^q$ terms in the fully connected (all-to-all) SYK model. Despite the exponentially large Hilbert space dimension, $2^N$, the sparse SYK model can be efficiently simulated using  Krylov subspace methods and massive parallelization. The results presented here are the first steps in a program to study finite $N$ corrections in the holographic correspondence between SYK and $AdS_2$ gravity with the goal of understanding how the  gravitational behavior of a traversable wormhole emerges from the low energy limit of two coupled SYKs as $N$ increases. 

Several known results from hypergraph theory provide us a good intuition about the effect of the sparsity in the model. We have studied the algebraic entropy, the spectral gap and the vertex expansion, which characterize the expansion in a hypergraph. These measures suggest that the $q$-body SYK model can be well approximated by a sparse model on a $(k\,q, q)$-regular hypergraph with a sparsity parameter $k$ of $\mathcal{O}(1)$. We find that the precise value of $k$ can be smaller for larger $q$. The introduction of sparsity allowed us to explore $q$-interacting Hamiltonians with $q=4$ \emph{and} $q=8$. The case of $q=8$ has been less explored in the literature \cite{Xu:2020shn, Alet:2020ehp} because it is computationally more demanding than $q=4$ since the number of terms in the Hamiltonian scales as $\sim N^q$. For concreteness, we  fixed the sparsity parameter to be $k=4$ and $k=2$ for  $q=4$ and $q=8$ cases respectively.

We showed that the two coupled sparse SYK model displays similar features as the corresponding all-to-all model such as a ground state that is close to a TFD and the presence of an energy gap. Finite $N$ effects are stronger at weak coupling where the system fails to reproduce the expected energy gap scaling obtained from the large $N$, and gravitational, analysis. Interestingly, the  scaling $\propto \mu^\frac{1}{2-2/q}$ predicted by holography is observed at finite $N$ only for $q=8$ for some interval of the coupling strength. 

The analysis at very small coupling $\mu$ is tricky. For small couplings, below $\mu\simeq 0.15$ for $q=4$ and below $\mu\simeq 0.015$ for $q=8$, we find that the overlap between the TFD and the ground state \eqref{eq:overlap} slightly deviates from unity, and the corresponding effective temperature of the TFD state deviates significantly from the analytical formula \eqref{eq:beta-mu}. This suggests  the existence of strong finite-size effects in the weak coupling regime. This is also the same range of $\mu$ where the energy gap obtained in Section \ref{subsec:Egap} does not follow the expected scaling $\propto \mu^\frac{1}{2-2/q}$ associated to the wormhole phase.

We also studied the transmission of signals between the two coupled sparse SYKs. At zero temperature and for $q=8$, we found a range of coupling strength that agrees with the prediction from holography, which is approximately the same range where the energy gap obeys the scaling \eqref{eq:Egap}. This might be an indication that finite-size effects are suppressed by additional $1/q$ corrections. Thus, it would be interesting to push the simulation at even larger $q$, which is still doable for the sparse model. Finally, we have also studied the effect of temperature on the two-point functions and the transmission amplitude. We found similarities with the phase transition between a wormhole and two high temperature black holes \cite{Maldacena:2018lmt}. These results are the first steps towards understanding the complete phase diagram of the coupled sparse SYK model. This is a  question that deserves further investigation, specially in the region of small coupling where finite-size effects are dominant. 

Our results underscore the significance of the sparse SYK as a holographic dual and highlights the need to better understand this class of models. There are several directions that can contribute to this aim. 

\paragraph{More properties of regular hypergraphs:} 
Hypergraphs that are optimal expanders, in the sense that their spectral gap is as large as possible, are known as Ramanujan hypergraphs. A $(kq, q)$-regular hypergraph is Ramanujan if any eigenvalue $\lambda_i$ of the adjacency matrix $A$ such that $\lambda_i \ne \lambda_\text{max}=k\,q(q-1)$, satisfies $|\lambda_i -q +2|\le 2 \, \sqrt{(k\, q-1)(q-1)}$. In \cite{Dumitriu_2019} the authors showed that $(k\,q, q)$-regular hypergraphs are \emph{almost} Ramanujan, \emph{i.e.} they satisfy
\begin{equation}\label{eq:Raj_theorem}
    |\lambda_i -q +2|\le 2 \, \sqrt{(k\,q-1)(q-1)} +\epsilon_N,
\end{equation}
where $\epsilon_N \rightarrow 0$ as $N\rightarrow \infty$. The sparse SYK model defined on $( k\, q, q)$-regular hypergraphs then satisfy \eqref{eq:Raj_theorem} in the $N\rightarrow \infty$ limit. But numerical studies of sparse SYK are carried out at finite $N.$ Thus, it would be interesting to investigate to what extent $( k\, q, q)$-regular hypergraphs satisfy \eqref{eq:Raj_theorem} for \emph{finite} $N.$  This could provide another indication of the  optimal sparsity for a given $N$ and $q$. Another property of uniform, regular hypergraphs is  how the `cover time' in a random walk scales with $N$. The vertex cover time, the edge cover time, and the inform time, they all scale like $N\,\log N$ when $N\rightarrow \infty$ \cite{Cooper_2013}. It would be interesting to understand the relevance of these cover times at finite $N$ in the context of a sparse SYK model. 

\paragraph{OTOCs:} The estimation that quantum chaos happens for $k\gtrsim 1$ derived in Ref.\,\cite{Garcia-Garcia:2020cdo} used of level statistics analysis. Another avenue to study aspects of quantum chaos is to compute out-of-time order correlators (OTOCs). One interesting direction is to study OTOCs in the single sparse SYK  using the numerical approach of Ref.\,\cite{Kobrin:2020xms}.  In the context of the  two coupled sparse SYK model studied in the present work, one can investigate the behavior of  the   Lyapunov exponent in the large $N$ limit.  In  \cite{Nosaka:2020nuk} the authors found that, for the all-to-all coupled SYKs, the high temperature phase saturates the chaos bound $2\pi/\beta$, while the low temperature phase has a small but non-zero Lyapunov exponent. The chaos exponent was related to the energy gap $E_\text{gap}$ via $\lambda_L\sim \exp(-(\frac{q}{4}-1)\beta E_\text{gap})$. It would be interesting to see whether this relationship still holds or how it is modified at finite $N$.

\paragraph{Quantum complexity:} Another question about the two coupled SYK model is how doable it is to simulate such system in a near-term quantum computer \cite{Garcia-Alvarez:2016wem}. One could investigate if the addition of sparseness makes the simulation on a quantum device easier and estimate the complexity to simulate it using methods such as qubitization \cite{Babbush:2018mlj, Xu:2020shn}. There is also the problem of encoding the state on a quantum device. The Majorana fermions can be encoded into qubits via the Jordan-Wigner transformation, but other choices of fermionic encoding \cite{Bravyi-Kitaev, Chien:2020sjq} could be applied to make the simulation more efficient. Going even further, one could also think of optimizing the simulation of SYK in hybrid quantum-classical format, e.g., by using tensor networks \cite{Yuan:2020xmq}.

\paragraph{Connection with teleportation protocols:} Quantum teleportation has been investigated in two coupled SYK systems \cite{Gao:2019nyj}, and in more generality for many-body quantum system \cite{Schuster:2021uvg, Brown:2019hmk, Nezami:2021yaq}. The key difference as compared to the eternal traversable wormhole setup of Maldacena and Qi is that the interaction is turned on only for a finite amount of time. A natural question in how to relate these teleportation protocols to the two coupled SYK system studied here. Another natural question is to quantify the amount of information that can be transferred through it. From the bulk perspective, this can be done by studying the effect of the backreaction of the signal that we want to send \cite{Maldacena:2017axo, Freivogel:2019whb, Caceres:2018ehr}. If the backreaction gets too strong the geometry will be deformed in such a way that it makes the opening of the wormhole smaller. From the quantum teleportation point of view, the bound on information transfer can be quantified from the amount of entanglement that is available and is understood as entanglement being used as a resource for the success of the teleportation.

We hope to return to some of these ideas in the near future. 


\section*{Acknowledgements}

It is a pleasure to thank Jason Pollack and Amir Raz for useful discussions. We also thank Antonio Garc\'ia-Garc\'ia, Dario Rosa and Brian Swingle for reading the draft and providing comments.  EC and RP are supported by National Science Foundation (NSF) Grant No. PHY-2112725. AM is supported by  NSF Grant No. PHY-1914679 and by a University Graduate Continuing Fellowship. The authors acknowledge the Texas Advanced Computing Center (TACC) at The University of Texas at Austin for providing HPC resources that have contributed to the research results reported within this paper. 


\appendix

\section{Numerical methods}
\label{app:numerical}

\subsection{Krylov subspace methods}

Krylov subspace methods are particularly useful when dealing with large sparse matrices $M$ whose action on a vector $x$ is easy to compute. These methods do not require writing down the full matrix explicitly, thus they significantly reduce the memory usage. 

Given a matrix $M$ and a vector $x$, the \textit{Krylov subspace} is define as the vector space
\begin{equation}
    \mathcal{K}_m(M,x) = \text{span}\{x,Mx, M^2x, \ldots, M^{m-1}x\}.
\end{equation}
Within the Krylov subspace, we can look for good approximations for eigenvectors. As $m$ becomes larger, the better the approximation, but as a trade-off it requires more storage. A \textit{Krylov process} consists in finding a good basis for the Krylov subspace. The vectors $x, Mx, M^2x,\ldots$ in general do not provide a good basis because $M^mx$ points more and more at the direction of dominant eigenvectors as $m$ increases. We can find a better basis, for example, by choosing $x$ randomly first, and making it orthonormal by using the Gram-Schmidt method. There are other good basis constructions that are not necessarily orthonormal. Two widely used Krylov processes are: \textit{Arnoldi} and \textit{unsymmetric Lanczos} process. In the Arnoldi process the basis is orthonormal.

Krylov subspace can be used to approximate the time evolution of a many-body quantum system. The comments below are based on section 5 of \cite{Luitz:2017yps}. 
We consider a quantum many-body system with Hamiltonian $H$ and Hilbert space dimension $\text{dim}\mathcal{H}$. The quantum state at time $t+\Delta t$ can be well approximated by a vector in the $m$-dimensional Krylov subspace
\begin{equation}
    \mathcal{K}_m = \text{span}\{|\psi(t)\rangle, H|\psi(t)\rangle, H^2|\psi(t)\rangle,\ldots,H^{m-1}|\psi(t)\rangle\}.
\end{equation}
We can generate an orthonormal basis for $\mathcal{K}_m$ using the Arnoldi process. Then, we project the Hamiltonian into this subspace to obtain an approximation for the time evolution
\begin{equation}
    e^{-iH\Delta t}|\psi(t)\rangle \simeq V_m e^{-i V_mHV_m\Delta t}e_1,
\end{equation}
where $V_m\in\mathbb{C}^{\text{dim}\mathcal{H}\times m}$ is the matrix whose columns contain the orthonormal basis vector of $\mathcal{K}_m$, and $e_1\in \mathbb{C}^m$ is the first unit vector of the basis, which corresponds to $|\psi(t)\rangle$ in the new basis.

\subsection{Jordan-Wigner transformation}

The SYK system with $N$ Majorana fermions can be encoded into spin-$\frac{1}{2}$ degrees of freedom via the Jordan-Wigner transformation. For Majorana fermions $\{\chi^j\}_{j=1}^N$ the transformation is
\begin{equation}
    \chi^{2n-1} = \frac{1}{\sqrt{2}}\left(\prod_{j=1}^{n-1}\sigma_j^z\right)\sigma_n^x, \qquad 
    \chi^{2n} = \frac{1}{\sqrt{2}}\left(\prod_{j=1}^{n-1}\sigma_j^z\right)\sigma_n^y, \qquad \{\chi_i,\chi_j\} = \delta_{ij}.
\end{equation}
The index $j=1,....,N$ labels the $j$th Majorana fermion, and we splitted them into even ($i=2n$) and odd ($i=2n-1$) cases. There is some freedom to define the transformation, the condition we need to obey is that they satisfy the Dirac algebra $\{\chi_i,\chi_j\} = \delta_{ij}$. There are other ways to represent fermionic operators as qubits, such as the Bravyi-Kitaev transformation \cite{Bravyi-Kitaev} (see also \cite{Chien:2020sjq} for a recent discussion on fermionic encoding). 

The Hilbert space associated to the single SYK Hamiltonian has dimension $2^{N/2}$. In the two coupled system, we can apply the Jordan-Wigner transformation twice so that the coupled system gets encoded into $N$ spin-$\frac{1}{2}$ degrees of freedom, with a corresponding Hilbert space of dimension $2^N$.

\end{document}